# Unravelling Chemical Exchanges Through Steady State Free Precession NMR

Sundaresan Jayanthi, Adonis Lupulescu, Mark Shif, Zuzana Osifová, and Lucio Frydman*

Department of Chemical and Biological Physics, Weizmann Institute of Science, Rehovot, Israel

## Abstract

NMR is uniquely endowed to analyze dynamics, with line shape and relaxation measurements covering timescales over several orders of magnitude. Further insight arises from pulse sequences like chemical exchange saturation-transfer or relaxation dispersion, which facilitate, respectively, the detectability and shift characterization of even lowly populated states, and the pinpointing of the exact exchange rates. The present study demonstrates that Steady State Free Precession (SSFP) experiments involving a train of pulses with flip angle α spaced by repetition times TR, combine valuable features from both these experiments. Indeed, in the presence of chemical exchanges, SSFP yields via its offset-dependent excitation and saturation profiles, detailed information about the number, the chemical shifts and the populations of the exchanging sites –even when these involve multiple intermediates with dissimilar abundances. Simultaneously SSFP can provide, via its TR dependence, a controllable timescale yielding kinetic information over a variety of slow/intermediate/fast exchange rates. All this is theoretically demonstrated with the aid of a Liouville-space formalism examining the steady state in the presence of chemical exchange. This formalism leads in both the slow and fast exchange regimes to analytical predictions that match well "brute-force" numerical calculations, which lend themselves to rapid and accurate fittings of exchange rates, chemical shifts, and site populations. The basic features associated to this novel approach to examine chemical kinetics are experimentally verified on simple model compounds; potential extensions are briefly discussed.

*Corresponding author. Email: lucio.frydman@weizmann.ac.il

## 1. Introduction

Nuclear magnetic resonance (NMR) is commonly used to elucidate chemical structures of organic, pharmaceutical and biological molecules in liquid and solid phases.[1] NMR is also uniquely poised to probe the nature of dynamic processes affecting these molecules or materials, at an atomic level. Dynamic information was first inferred from the NMR relaxation times $T_1$ and $T_2$;[2] as spectral resolution improved it became clear that chemical interconversions would also alter the ensuing NMR line shapes.[3-5] In the simplest case of a single chemical site interconverting between two frequencies $\omega_A$ and $\omega_B$, McConnell's extension of the Bloch's formalism[6] showed that the ratio between the rate of exchange $k_{ex}$ and the chemical shift difference (in Hz) $|\omega_A-\omega_B|$ defines a "timescale". Depending on whether this $|\omega_A-\omega_B|/k_{ex}$ ratio would be smaller, equal or larger than one, kinetic changes would modify line shapes –from separate resonances undergoing excess exchange broadenings in proportion to $k_{ex}$, to single population-weighted peaks undergoing progressive line narrowing in proportion to $1/k_{ex}$. These line shape analyses were rapidly generalized to encompass multiple interactions, multiple exchanging sites, and arbitrary populations,[7-9] and have become standard practice for elucidating tautomerisms, chemical reactions and rotations, in liquid and solid systems.

Over time, certain limitations of the basic Bloch-McConnell line shape analyses, were noted. If for instance the exchange rate was either too slow to introduce a measurable peak broadening or too fast to leave a discernible excess line width, if the exchanging populations were too dissimilar, or if the linewidths of the exchanging peaks were very different, 1D line shape fits could become underdetermined. Equally hard to tackle were large macromolecules, where peak crowding and relatively short $T_2$ relaxation times limited the usefulness of 1D data. This stimulated proposals such as the saturation/inversion-transfer experiment, where a site is initially taken away from equilibrium and then allowed to perturb, via mutual chemical exchanges, the response of a second interconverting site.[10] These experiments –which saw extensions to variants like 2D Exchange Spectroscopy, 2D ZZ-exchange NMR, chemical exchange saturation transfer (CEST) and dark-state exchange saturation transfer– offered a dynamic insight exceeding traditional 1D line shape analyses.[11-13] Besides being able to probe larger molecules and slower timescales, selective saturation approaches provided an opportunity to magnify the effects of a slow exchange by factors $\approx k_{ex}*T_1$, thereby serving to highlight populations that might otherwise be too small to be detected. On the opposite, fast-exchange regime, so-called relaxation-dispersion experiments were developed based on the Carr-Purcell Meiboom-Gill (CPMG) sequence. By contrast to inversion-based methods, CPMG imposes a controllable timescale for probing the exchange rates, based on the delay between the refocusing pulses it applies. Exchange processes happening on such timescales will contribute to an additional transverse dephasing, enabling rates to be measured.[9,14] This, however, is done under a set of limitations –foremost the fact that it is generally feasible to fit such relaxation-dispersion (RD) data by a two-site exchange model with suitably chosen rates, populations and offsets –thereby masking more complicated multi-site exchange process and/or being vulnerable to ambiguities in the fit. Pulse imperfections, scalar couplings, and differential relaxation of in- and anti-phase terms can also complicate analyses of these experiments. At some point, and despite their differences, CEST- and RD-related methodologies eventually intersect, as decreasing RD's interpulse delay to zero leads to a continuous spin-locking pulse akin to that used in saturation-transfer experiments. Both dynamics-oriented experiments then end up measuring $R_{1\rho}$ and $R_{2\rho}$ properties of the exchanging system, with effective radiofrequency (RF) fields and offsets chosen to highlight faster or slower processes as per the kind of problem being probed.[15,16]

It has been recently shown that Steady-State Free Precession (SSFP) also can, under suitable conditions, provide high resolution NMR lineshapes.[17,18] The SSFP pulse sequence (Figure 1a) consists of a long ($\geq T_1$) train of equally spaced RF pulses with constant flip angles $\alpha$, applied at repetition times $TR \ll T_2, T_1$. The phase of these pulses may be constant or incremented as $\phi_k = (k-1)\Delta\phi$, where $\phi_k$ is the phase of the $k^{th}$ pulse in the train and $\Delta\phi$ represents the phase increment. After a long enough train of pulses Carr showed that an isolated spin ensemble will then be driven into a steady state, with coexisting spin- and stimulated-echoes leading to constant transverse and longitudinal magnetizations.[19] SSFP's attractiveness derives from the fact that, if relaxation rates $R_2 = (T_2)^{-1}$ and $R_1 = (T_1)^{-1}$ are similar, the ensuing transverse magnetization becomes ca. 50% of the equilibrium magnetization $M_o$.[20,21] In addition, Carr showed that for nearly any flip angle used, the $R_2 \approx R_1$ condition prevalent in liquids will also lead to saturation "dark bands" where $M_z \approx 0$. If one were to consider the examination of a chemical exchange process by SSFP, this would mean that on one hand the sequence's *TR* can provide a "timescale" that is similar to that endowed by interpulse delays in CPMG experiments; while the sequence's "dark bands" can provide a likeness to the saturation frequencies arising when combining CEST with multiple-pulse experiments.[22] The latter feature has in fact been noticed in CEST MRI, where suitable *TR*s and offsets of a balanced SSFP experiment were used to collect a water-based image while achieving a selective saturation of the choline, glucose or glycogen resonances.[23] Such bSSFPx experiment was then analyzed in the slow-exchange and low RF saturation regimes typical of CEST MRI based on an $R_{1\rho}/R_{2\rho}$ formalism.[15,16]

The present study revisits the problem of chemical exchange and SSFP, but for arbitrary rates, populations and number of exchanging sites. It is thus found that SSFP ends up possessing several of the features conveyed by CEST and RD experiments, while helping to overcome each other's respective limitations: As in RD, the reliance of SSFP on periodic pulsing endows it with the capability of pinpointing the rates of the chemical exchange – including the possibility of relying on analytical expressions in the slow and fast exchange regime. And as in CEST, the presence of controllable saturation bands endows SSFP with the ability of reading out the number and the chemical shifts of the exchanging sites in a model-free fashion, while amplifying the effects of lowly populated sites. The following Sections introduce the model used to interpret the SSFP experiment in chemically exchanging systems, present examples of its main features, describe experimental verifications, and puts forward a pipeline for its practical use.

## 2. SSFP and Chemical Exchange: Theoretical Treatment

The transverse and longitudinal magnetizations of an SSFP experiment implemented on an isolated spin-1/2 ensemble will depend on relaxation rates, on the site's offset $\Delta\omega$, and on flip angle $\alpha$.[19-21] These magnetization components will repeat themselves at regular $1/TR$ intervals (Figure 1b), as is reflected by the analytical expressions for the transverse

$$M_{SS,+}(\Delta\omega) = M_{SS,x}(\Delta\omega) + iM_{SS,y}(\Delta\omega) = M_0 e^{\frac{i\pi}{2}} \frac{(a\exp(i\Delta\omega TR) + b)}{\mathrm{c}\cdot\cos(\Delta\omega TR) + d} \qquad (1a)$$

and longitudinal

$$M_{SS,z}(\Delta\omega) = M_0 \frac{e\cdot\cos(\Delta\omega TR) + f}{c\cdot\cos(\Delta\omega TR) + d} \qquad (1b)$$

magnetizations.[24] The $a, b, c, d, e, f$ coefficients in Eq. [1] depend on relaxation factors $E_1 = \exp(-R_1 TR)$, $E_2 = \exp(-R_2 TR)$, and are: $a = -E_2(1-E_1)\mathrm{S}, b = (1-E_1)\mathrm{S}, c = -E_2(1-E_1)(1+\mathrm{C}), d = 1 - E_1 cos\alpha - {E_2}^2(E_1 - C), e = -E_2(1-E_1)(1+cos\,\alpha), f = (1-E_1)(E_2^2 + cos\,\alpha)$, with $C = \cos\alpha, S = \sin\alpha$. The $\boldsymbol{M}_{SS}(\Delta\omega)$ vector in Eq. [1] reflects the steady-state magnetization immediately after any pulse, and clearly fulfils $\boldsymbol{M_{SS}}(\Delta\omega) = \boldsymbol{M_{SS}}(\Delta\omega + 2\pi n/TR), n$ integer. Notice that offsets leading to minimal transverse magnetizations, coincide with those leading to maximum longitudinal saturation (Fig. 1b).

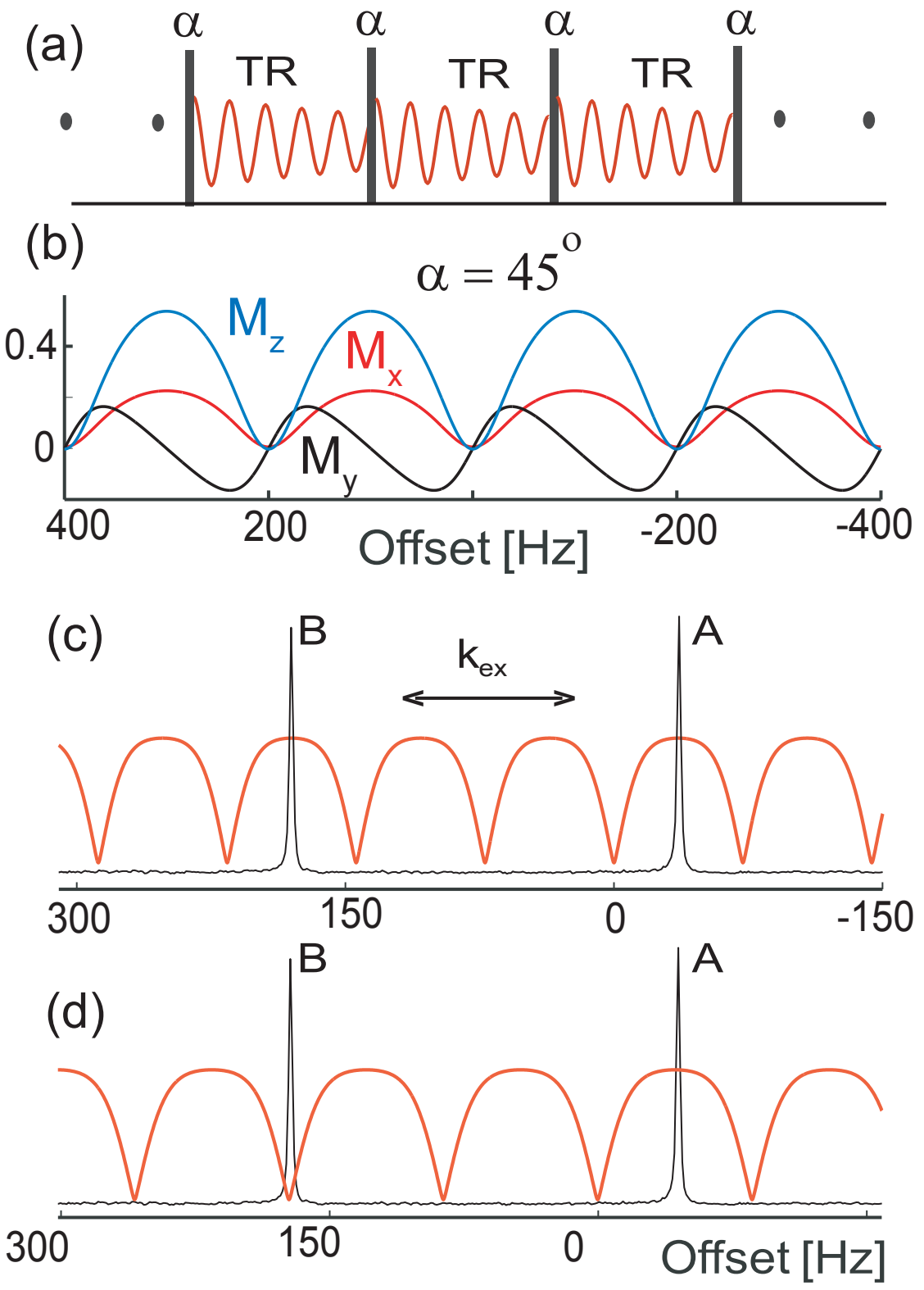

**Figure 1.** (a) SSFP pulse sequence consisting of pulses of flip angle α, separated by periods of free evolution TR. (b) Periodic behavior of the steady state magnetization $\boldsymbol{M}_{SS} = (M_{SS,x}, M_{SS,y}, M_{SS,z})$ for a single site, calculated for $\alpha = 45^\circ$, $TR$ = 5ms, $\Delta\phi = 0$, and for offsets extending over a few multiples of 1/TR. These profiles are plotted as function of offset for a constant pulse phase, but will be equivalent to the profiles arising as function of phase increment $\Delta\phi$ with a $\Delta\omega = 0$: Phase increments within the $[-\pi\ \pi]$ interval correspond to incrementing offsets $\Delta\omega$ in the $[-\pi/TR\ , \pi/TR]$ interval. (c,d) SSFP profiles for two different $TR$'s, superimposed on idealized spectra of a two-site system in slow exchange. Relaxation rates considered in (b-d) were $R_1 = 1\ \text{s}^{-1}, R_2 = 5\ \text{s}^{-1}$; thermal magnetization was $M_0 = 1$.

Consider now two sites with different chemical shifts, being subject to such experiments. If sites are irradiated with an SSFP profile like that shown in Fig. 1c, both would attain maximal magnetization; if irradiated with the profile in Fig. 1d, site $B$ will reach saturation while site $A$ remains unchanged. However, if there is chemical exchange between the two sites, one would expect the saturation of site $B$ to induce a drop in the magnetization of site $A$, which will depend on the populations and rates involved in the exchange. This modulation of one site $A$ by another site $B$ constitutes one of the SSFP signatures of chemical exchange that are here pursued. Notice that changing the SSFP profile from that in Fig. 1c to that in Fig. 1d entails changing the TR while maintaining a constant offset (for instance, one that maximizes the SSFP profile of site $B$); such variable-TR approach was the method chosen to probe these phenomena.

For describing this process we consider first a classical two-site exchange, involving sites $A$ and $B$ of populations $p_A$ and $p_B$, and chemical shifts $\omega_A$, $\omega_B$. The exchange rate $k_{ex} = k_{AB} + k_{BA}$ characterizing the process is the sum of the exchange rates $k_{AB}$ (from $B$ to $A$) and $k_{BA}$ (from $A$ to $B$), and at thermal equilibrium $k_{BA}p_A = k_{AB}p_B$. Denoting by $M_i^A(t)$ and $M_i^B(t)$ $\{i = x, y, z\}$ the magnetizations of sites $A$ and $B$ during free evolution, these can be determined by solving the homogeneous system of equations

$$d\boldsymbol{\mathcal{M}}/dt = -\mathcal{L}\boldsymbol{\mathcal{M}}(t)\,, \qquad (2a)$$

where

$$\boldsymbol{\mathcal{M}}(t) = \begin{pmatrix}1/2 & M_x^A(t) & M_y^A(t) & M_z^A(t) & M_x^B(t) & M_y^B(t) & M_z^B(t)\end{pmatrix}^T \qquad (2b)$$

and

$$\mathcal{L} = \begin{pmatrix} 0 & 0 & 0 & 0 & 0 & 0 & 0 \\ 0 & R_{2A}+k_{BA} & \omega_A & 0 & -k_{AB} & 0 & 0 \\ 0 & -\omega_A & R_{2A}+k_{BA} & 0 & 0 & -k_{AB} & 0 \\ -2R_{1A}M_A^0 & 0 & 0 & R_{1A}+k_{BA} & 0 & 0 & -k_{AB} \\ 0 & -k_{BA} & 0 & 0 & R_{2B}+k_{AB} & \omega_B & 0 \\ 0 & 0 & -k_{BA} & 0 & -\omega_B & R_{2B}+k_{AB} & 0 \\ -2R_{1B}M_B^0 & 0 & 0 & -k_{BA} & 0 & 0 & R_{1B}+k_{AB} \end{pmatrix}. \quad (2c)$$

Here $\{R_{1i}, R_{2i}\}_{i=A,B}$ represent the longitudinal and transverse relaxation rates [in $s^{-1}$] of the sites, and since the thermal magnetizations $M_A^0$ and $M_B^0$ are proportional to their respective populations $p_A$ and $p_B$, we take for simplicity $M_A^0 = p_A, M_B^0 = p_B$.

The system's propagator over a period $TR$ will be given by $U_{\text{free}}(TR, 0) = \exp(-\mathcal{L}TR)$. The steady state condition implies that the ensemble magnetization immediately after $n^{th}$ and $(n+1)^{th}$ pulses, must be equal. This is akin to what is done when dealing with non-exchanging systems;[20,21,24] denoting $U_{RF}$ the propagator corresponding to the evolution imparted by the SSFP pulses, this demand leads to the equation

$$\boldsymbol{\mathcal{M}}_{SS}(n+1) = U_{RF}U_{\text{free}}\boldsymbol{\mathcal{M}}_{SS}(n) = \boldsymbol{\mathcal{M}}_{SS}(n) \qquad (3)$$

where $\boldsymbol{\mathcal{M}}_{SS}$ reflects the classical magnetizations of the two-site exchanging system. $\boldsymbol{\mathcal{M}}_{SS}$ can always be calculated numerically by increasing *n* and examining the steady states that are reached by each magnetization component after a sufficiently large number of pulses. Analytical expressions of $\boldsymbol{\mathcal{M}}_{SS}$ may also be obtained, provided an analytical expression of $U_{\text{free}}$ is available. This can be obtained under slow- or fast-exchange conditions, as treated below.

**2.1 SSFP in the Slow Two-Site Exchange Regime.** Slowly exchanging systems are characterised by $k_{ex} \ll |\omega_A - \omega_B|$. In such cases, the relative precession rate imparted by $\omega_A - \omega_B$ will exceed the changes introduced by $k_{AB}$ or $k_{BA}$; the effects of the off-diagonal transverse exchange elements in Eq. [3], can consequently be neglected.[25] This approximation allowed us to derive full analytical expressions for the SSFP magnetizations. These are given in Section 1 of the Supporting Information (SI) for the case of one of the sites being on-resonance, and for both sites having the same $R_1$ and $R_2$ relaxation rates. The ensuing *TR*-dependent SSFP profile $M_{SS}(TR)$ is explored further in Figures 2a-2c, as function of populations, exchange rates and flip angles. These numerical simulations focused on one of the exchanging sites (*A*) which was assumed fixed at an offset $\pi/TR$; in the absence of chemical

exchange this condition would provide maximal transverse SSFP magnetization (cf. Fig. 1c). The sensitivity of the ensuing $M^A_{SS,+}(TR)$ profiles to the exchange parameters is clear, as graphs show that even highly imbalanced populations and slow exchange processes –both strengths from CEST– will be reflected in these profiles. Notice as well the strong response of these profiles to the exchange rate, as well as their flip-angle sensitivity. Alongside these numerical SSFP profiles, arising from propagating along the SSFP pulse train for several times the relaxation time $T_1$ and then examining the ensuing steady state, Figures 2d-2f present $M^A_{SS,+}(TR)$ SSFP profiles predicted by the approximate analytical derivations. Except for the responses at short *TR*s, there is good agreement between the analytical and "brute force" steady-state predictions vis-à-vis all parameters. The deviations observed at short *TR*s indicate that the decays imposed by transverse off-diagonal exchange matrix elements can no longer be neglected in this regime vs the relative precession angle $(\omega_A - \omega_B)TR$ between the two magnetizations –hence breaking the simplification underlying the analytical derivation. Still, the good agreement between the "brute-force" numerical model and the analytical expressions, imply that the latter can be used as starting points for estimating exchange rates and/or populations from variable-$TR$ SSFP data.

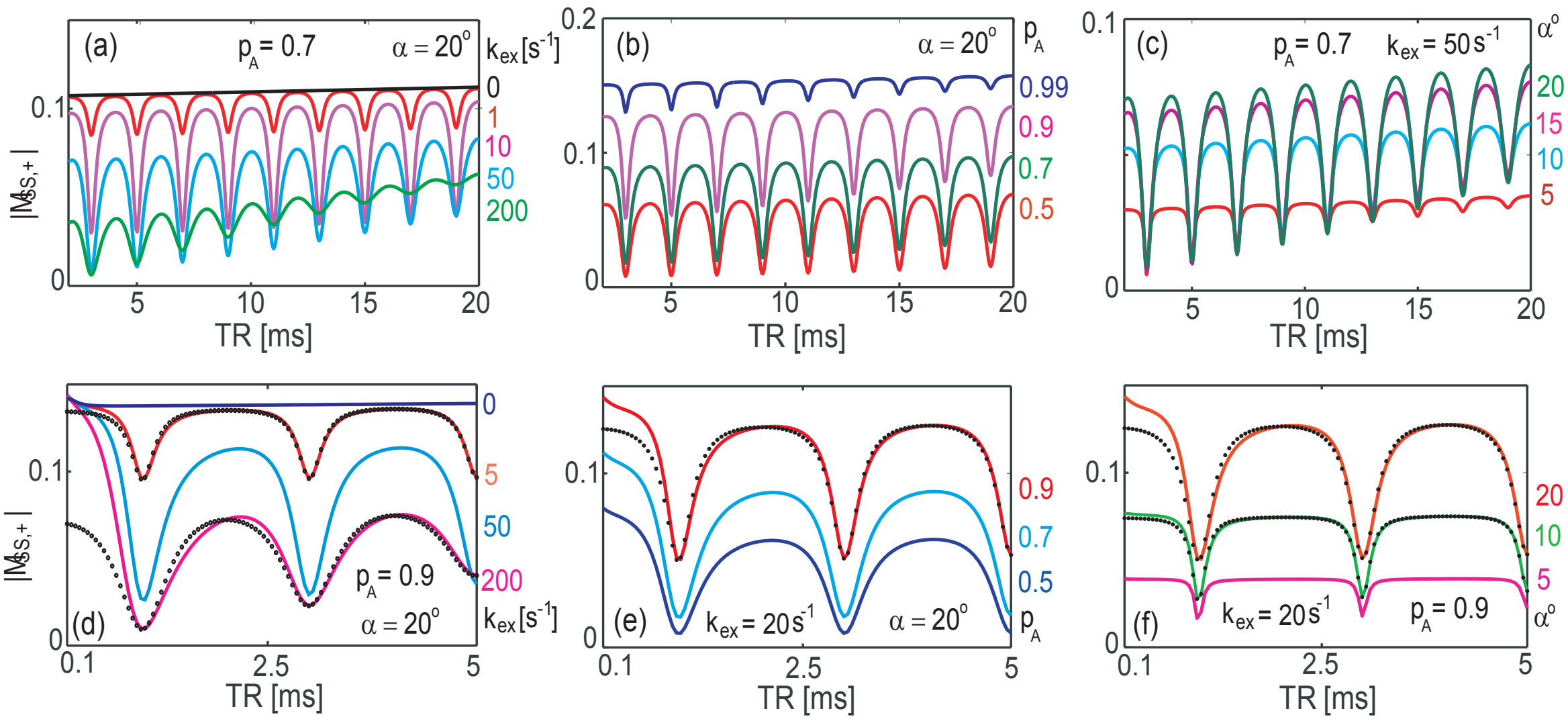

**Figure 2.** (a-c) SSFP profiles $\left|M^A_{SS,+}(TR)\right|$ computed for different rates, populations and flip angles, for a two-site $A \Leftrightarrow B$ exchange process where only the magnetization of site $A$ is being monitored. Identical relaxation rates were considered for the two sites ($R_1 = 1\ \mathrm{s}^{-1}, R_2 = 5\ \mathrm{s}^{-1}$); their chemical shift difference was $(\omega_B - \omega_A)/2\pi = 500$ Hz, and thermal magnetizations were assumed equal to the respective populations. (d-f) Overlays of approximate analytical steady-state solutions (dotted) against "brute force" SSFP profiles (continuous lines) for the indicated parameters. To implement such "on-resonance" measurements one could set the offset of site $A$ to zero and phase increment the pulses by $\Delta\phi = \pi$, or set $\omega_A = \frac{\pi}{TR}$ and $\Delta\phi = 0$.

In more general cases where relaxation rates are unequal ($R_{1A} \neq R_{1B}, R_{2A} \neq R_{2B}$) and/or offsets are arbitrary, the aforementioned $k_{ex} \ll |\omega_A - \omega_B|$ approximation still enables an analytical derivation of the steady state magnetizations. Section 2 of the SI presents part of the ensuing analytical solutions; unfortunately, these expressions are too long and unwieldy to provide much useful insight by their inspection. For the case of equal relaxation rates, results arising from these SSFP predictions upon placing the carrier at arbitrary frequencies, are displayed in Figure 3 for a variety of carrier frequencies and exchange rates. These curves show a complex behaviour as function of TR, which is clearly distinct from what will be observed in the absence of chemical exchange (blue curves). Also shown by continuous red curves in Figure 3 are numerical "brute force" solutions performed for identical parameters as those used with the analytical expressions; once again, except at short *TR*s, the agreement between the analytical and numerical steady-state solutions is very good. Also in this case we found that analytical expressions, though cumbersome, were useful in the iterative search of exchange parameters that best fit an experimental set.

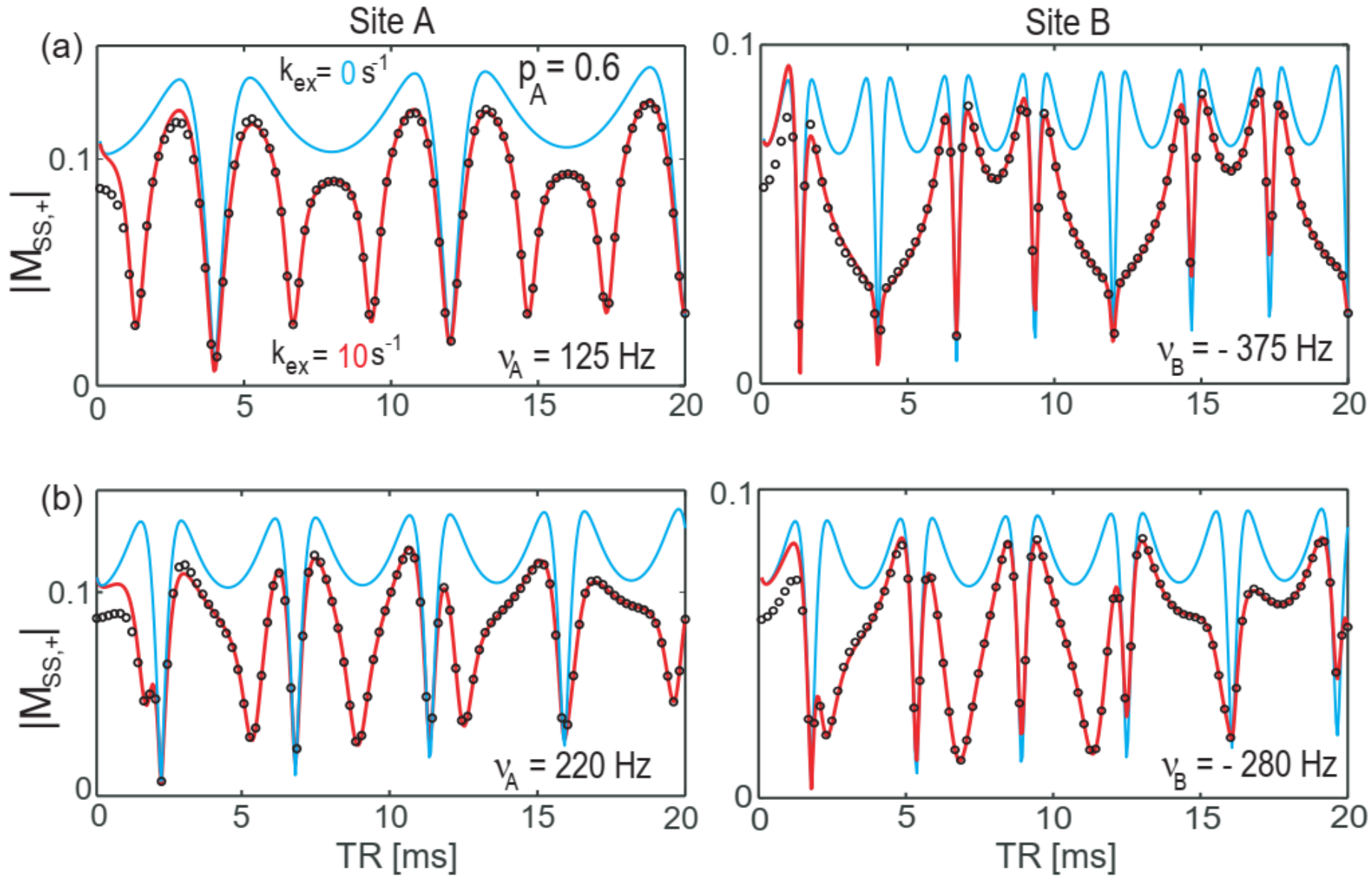

**Figure 3.** (a, b) SSFP profiles $|\mathcal{M}_{SS,+}(TR)|$ expected to arise from the sequence in Fig. 1a, applied on sites A and B separated by $(\omega_B - \omega_A)/2\pi = 500$ Hz for different carrier frequencies in the presence of chemical exchange (to appreciate these effects, $k_{ex} = 0$ profiles are also provided as blue traces). Simulations considered $R_{1A} = R_{1B} = 1\ \text{s}^{-1}, R_{2A} = R_{2B} = 5\ \text{s}^{-1}$, $\alpha = 23°$, constant phase (x) for the pulses, and thermal magnetizations equal to the respective sites' populations. Steady-state analytical solutions are shown as circles; "brute-force" numerical estimations as overlaid red lines.

**2.2 Two-Site SSFP: From the Slow to the Fast Exchange Regimes.** The preceding paragraph validated the arguments foretold in Figure 1 concerning SSFP's kinetic sensitivity in the slow-exchange regime, even when dealing with unequal populations. Figure 4 extends some of those two-sites considerations, with numerical "brute force" simulations presenting SSFP profiles for chemical exchange rates ranging from the slow through the intermediate and on to the fast kinetic regime. Since in the intermediate and fast exchanges cases individual magnetizations of the two sites are not measurable, Figure 4a displays the total magnetization $M^T_{SS,+}(TR) = M^A_{SS,+}(TR) + M^B_{SS,+}(TR)$ for all rates. Shown in Figure 4b are the corresponding two-site exchange 1D NMR spectra, scaled by the indicated factors for easier visualization. Features that are clear from these simulations are the non-monotonic oscillations that the second site ("*B*") imparts in the slow exchange regime, the blurring of these inter-site oscillations in the intermediate regime, and their reappearance at an average offset in the fast exchange regime. Also interesting to note is the strong *TR* dependence that the exchange will have on the SSFP signals: these will be minimally affected by the exchange for very short *TR*s or for long *TR*s fulfilling SSFP's optimal offset conditions ($[\omega_{center} = (2k+1)\pi/TR]_{k=0,1\ldots}$), but drop to low levels for intermediate inter-pulse delays values fulfilling $TR \approx k_{ex}^{-1}$. Such "V" shape drop in SSFP magnetization parallels the drop in intensity that the 1D NMR data show vs rate (in SSFP, however, such drop can be modulated by changing the separation between the pulses –notice the nearly constant amplitude displayed in Fig. 4a for short *TR*s).

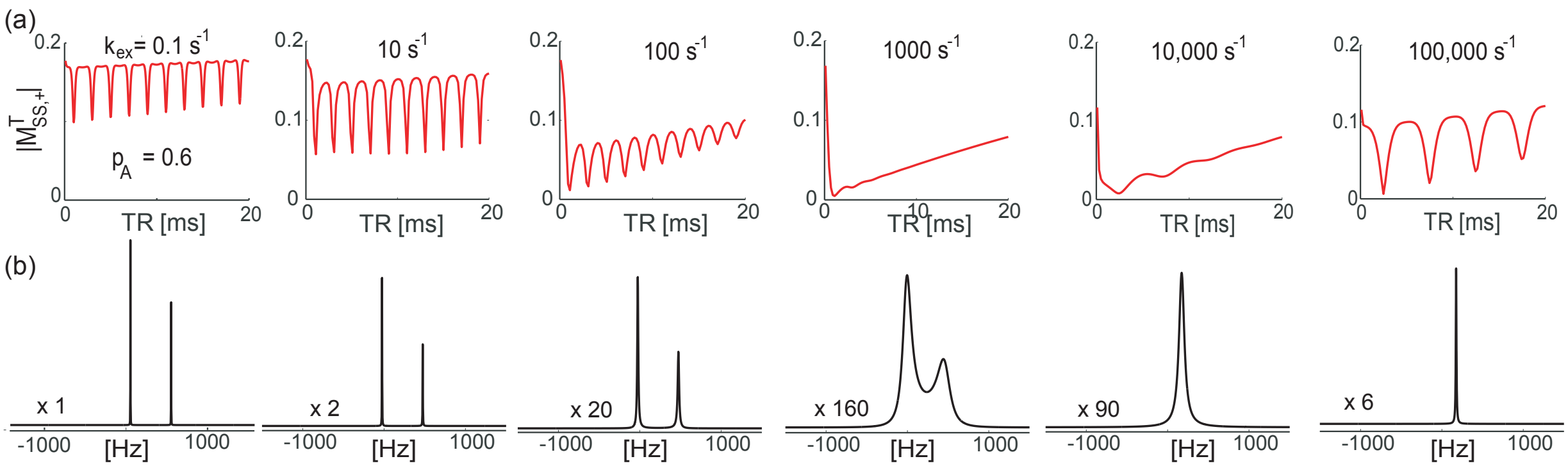

**Figure 4.** (a) SSFP profiles showing $|M^T_{SS,+}(TR)|$ for different chemical exchange rates. (b) Corresponding two-site exchange spectra. The difference of chemical shifts of sites A and B was assumed 500 Hz and $R_{1A} = R_{1B} = 1\ \text{s}^{-1}, R_{2A} = R_{2B} = 5\ \text{s}^{-1}, \alpha = 23°, p_A = 0.6$. All simulations assumed $\omega_A = 2\pi/TR$.

While numerical simulations are informative, analytical descriptions could further facilitate the analysis of kinetic SSFP data. Although an exact analytical description of the SSFP response over the full kinetic regime was not forthcoming, it was possible to quantitatively describe the SSFP dynamics in the fast exchange regime ( $k_{ex} > |\omega_A - \omega_B|$). In this case only a single resonance at frequency $\bar{\Omega} = \omega_A p_A + \omega_B p_B$ will be observed, characterized by effective longitudinal and transverse relaxation rates

$$\bar{R}_1 = R_{1A}p_A + R_{1B}p_B, \tag{4a}$$

$$\bar{R}_2 = R_{2A}p_A + R_{2B}p_B + \frac{p_A p_B}{k_{ex}}(\omega_A - \omega_B)^2\left(1 - \frac{\tanh(k_{ex}TR)}{k_{ex}TR}\right), \tag{4b}$$

where the last term includes the correction derived in the context of CPMG experiments.[9,26] Under these assumptions, and in analogy with Eq. [1], the SSFP magnetization can be described as

$$M_{SS,+}^{T}(TR) = \bar{M}_X + i\bar{M}_Y = e^{\frac{i\pi}{2}} \frac{(\bar{a}\, e^{i\bar{\Omega}TR} + \bar{b})}{\bar{c}cos(\bar{\Omega}TR) + \bar{d}} \tag{5}$$

where $\bar{a} = -\bar{E}_2(1-\bar{E}_1)\mathrm{S}, \bar{b} = (1-\bar{E}_1)\mathrm{S}, \bar{c} = -\bar{E}_2(1-\bar{E}_1)(1+\mathrm{C}), \bar{d} = 1 - \bar{E}_1 cos\alpha - \bar{E}_2{}^2(\bar{E}_1 - C),\ C = \cos\alpha, S = \sin\alpha$, and $\bar{E}_1 = e^{-\bar{R}_1 TR}, \bar{E}_2 = e^{-\bar{R}_2 TR}$ denote average values for these relaxation-weighting functions.

Calculations of $|\mathcal{M}_{SS,+}^{T}(TR)|$ for a range of chemical exchange rates are illustrated in Figure 5. "Brute force" numerical results are here compared against the predictions of Eq. [5], showing a good match for fast exchanges under both on- and off-resonance situations. Once again these curves show nearly constant signal intensity values for very short *TR*s, transitioning into Carr-like SSFP profiles with increasingly longer *TR*s as $k_{ex}$ increases. Deviations between the predictions of Eq. [5] and exact numerical calculations again appear for short *TR*s ($TR \leq \frac{2}{k_{ex}}$). More useful than $|M_{SS,+}^{T}(TR)|$ itself to compute exchange rates, might be the derivatives $|d|M_{SS,+}^{T}(TR)|/d(TR)|$ displayed in Figure 5b: the maxima of these derivatives reflect the changing widths of the $|M_{SS,+}^{T}(TR)|$ dips, which together with their position appear to be more sensitive to variation of the exchange rate. Section 4 of the SI examines further these profiles' dependence on the relative populations, exchange rates, and flip angles. These simulations suggest that small flip angles that are not necessarily associated to SSFP's highest sensitivity might have higher sensitivity to rates, and that this sensitivity persists even when dealing with very skewed populations.

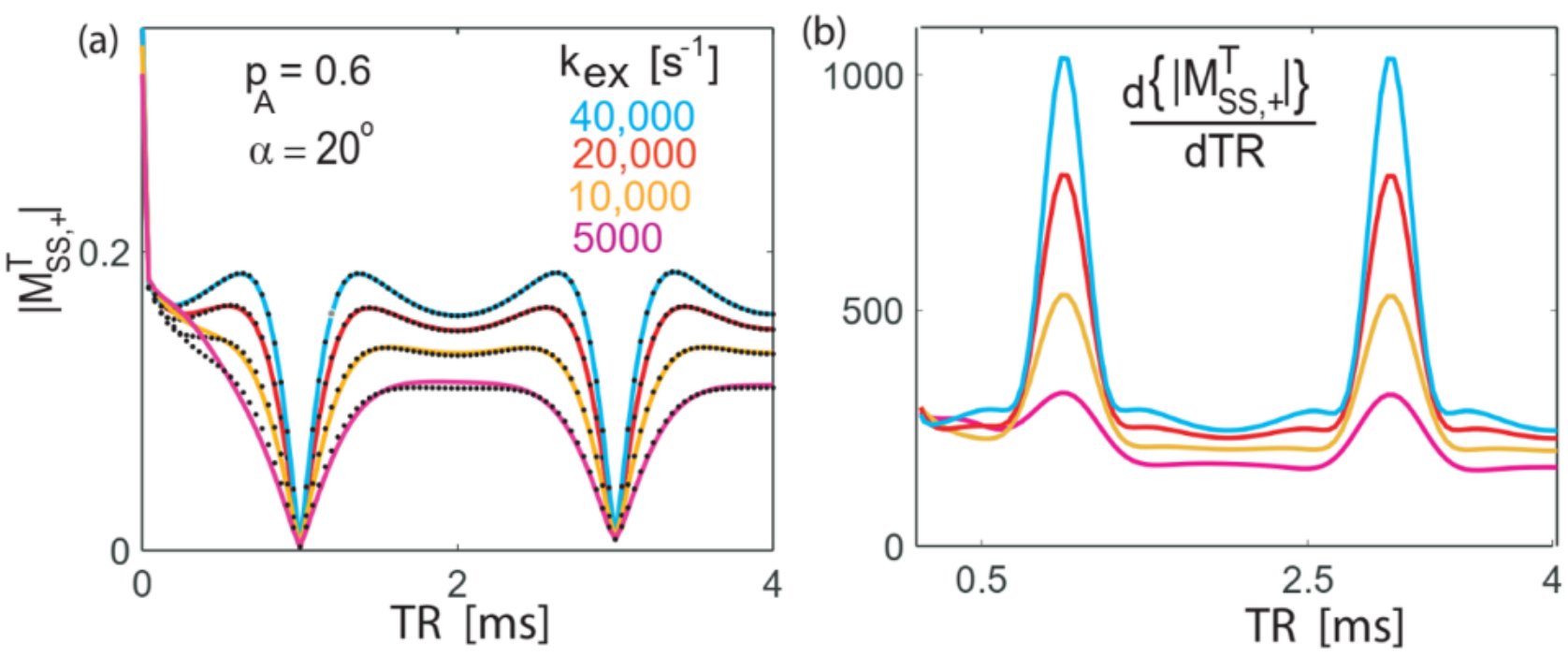

**Figure 5.** SSFP profiles of $|\mathcal{M}_{SS,+}^{T}(TR)|$ and of $|d|M_{SS,+}^{T}(TR)|/d(TR)|$ for different rates in the fast-exchange regime. The offset of the observed peak is 500 Hz; other parameters are $\alpha = 20°$, $R_{1A} = R_{1B} = 1\ \mathrm{s}^{-1}, R_{2A} = R_{2B} = 5\ \mathrm{s}^{-1}$.

**2.3 SSFP in the Presence of Multi-site Exchanges.** Based on the qualitative arguments in Figure 1 it follows that, in analogy with CEST, SSFP might serve to identify the number and chemical shifts of several exchanging sites. If considering, for example, a three-site exchange process, then the "on-resonance" ($\omega_A = 0, \Delta\phi = \pi$) variable-*TR* SSFP profile of site *A*, should exhibit dips whenever the $\mathcal{M}_{SS,z} \approx 0$ saturation condition coincides with the chemical shifts of either site B or site C. This qualitative picture is confirmed in Figure 6a, which for this simple on-resonance case enables distinguishing two-site from three-site chemical exchanges by mere visual inspection of $\mathcal{M}_{SS}^{A}(TR)$. In a more general case, the exact offset of the exchanging sites may be unknown and hence the SSFP patterns may become harder to analyse by simple visual inspection (Fig. 6b). However, the formalisms described above can be used to simultaneously fit these intricate patterns, and extract from them accurate kinetic, population, and chemical shift information even if involving multiple sites.

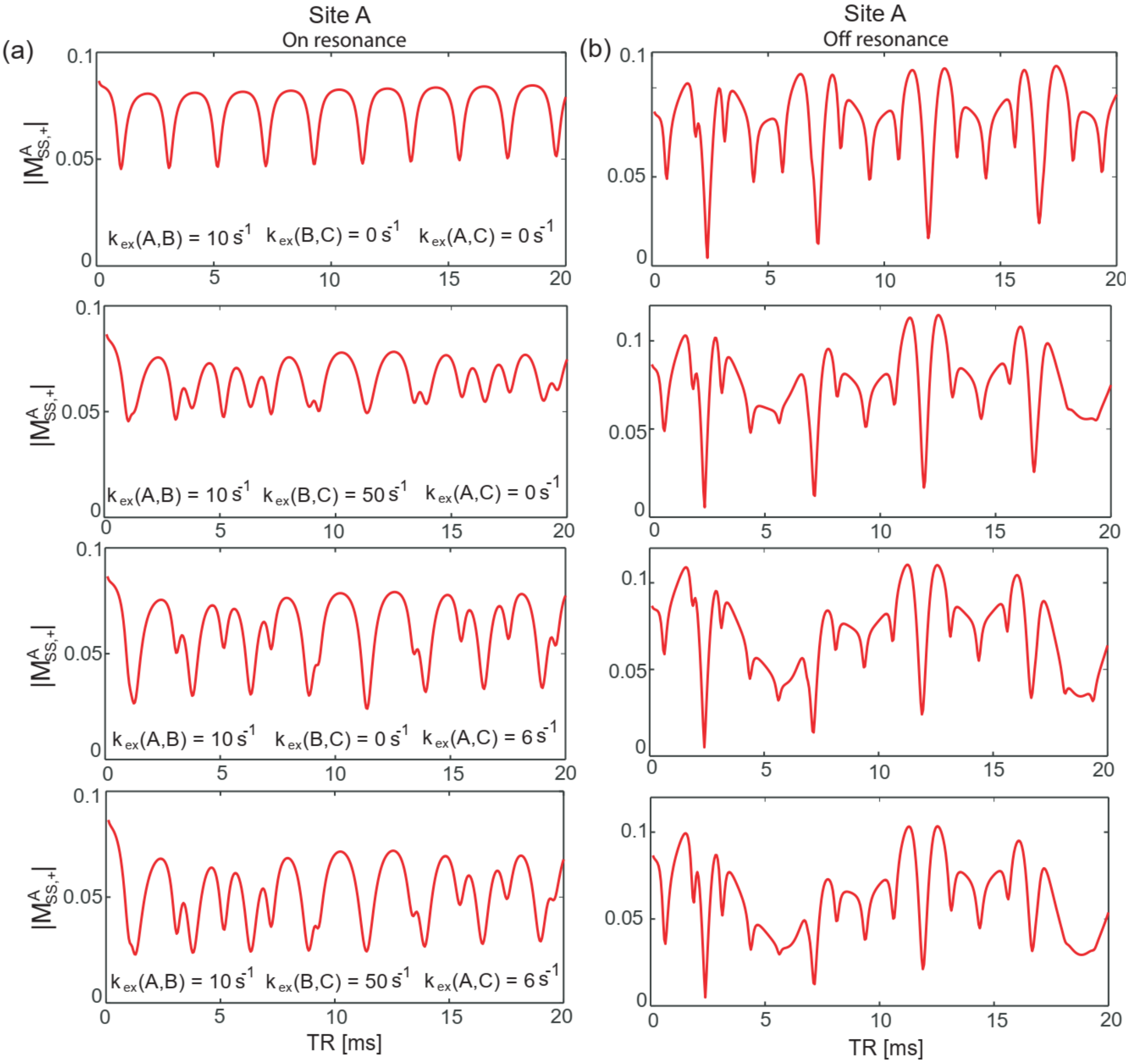

**Figure 6.** SSFP profiles for a three-site chemical exchange between sites A, B, and C. (a) On-resonance SSFP profiles for site A monitored as function of TR, for the indicated exchange cases. For this on-resonance site A case the offsets of sites B and C are 485 Hz and -395 Hz, respectively. (b) SSFP profiles of site A expected for the same chemical shift differences and kinetic rates as in panel (a), but upon moving the carrier 105 Hz. Populations were fixed at $p_A = 0.5, p_B = 0.1, p_C = 0.4$ and the thermal magnetizations of the three sites were taken equal to their respective populations. Other parameters used are $R_1 = 1\ \mathrm{s}^{-1}, R_2 = 5\ \mathrm{s}^{-1}$, $\alpha = 23°$ (common for all three sites).

## 3. Experimental

Variable-temperature NMR experiments were performed on a Bruker spectrometer equipped with a TCI Prodigy® probe, using an AVIII HD console and operated with Bruker's TopSpin 3.6.5 software. SSFP sequences were written based on a loop of equidistant pulses with flip angle $\alpha$ and alternating RF phases separated by delays $TR$ (Fig. 7a), followed by a conventional $\{^1H\}^{13}C$ NMR acquisition. This approach imparts an initial steady-state build-up, and is followed by acquisition with long acquisition times $T_{Acq}$ yielding good resolution and accurate signal intensities.[27] The number of looped pulses $n$ was varied in unison with TR to keep the $n{\cdot}TR$ product constant at $\approx 5{\cdot}T_1$; spectra thus reflected the $M_{SS}(TR)$ depicted above, while retaining full spectral resolution. Figures 7b and 7c show the chemical tautomerisms monitored in this study, involving two-site exchanges in N,N-dimethylacetamide (DMA)[28.29] and acetylacetone (AcAc).[30,31] Natural abundance $^{13}C$ was chosen for these SSFP tests to avoid potential complications from homonuclear NOEs or J-couplings that may arise with $^1$Hs. In both cases 50% vol. solutions were prepared by mixing 300 μL of the targeted molecules (both from SigmaAldrich) in 300 μL of deuterated solvent: $CDCl_3$ was used for DMA and DMSO-$d_6$ for AcAc. In the latter case, and to accelerate the chemical tautomerism, a diluted aqueous NaOH solution was added dropwise (10 μL at a time). After the addition of NaOH, the solution turned slightly yellow, indicating the formation of an enolate.

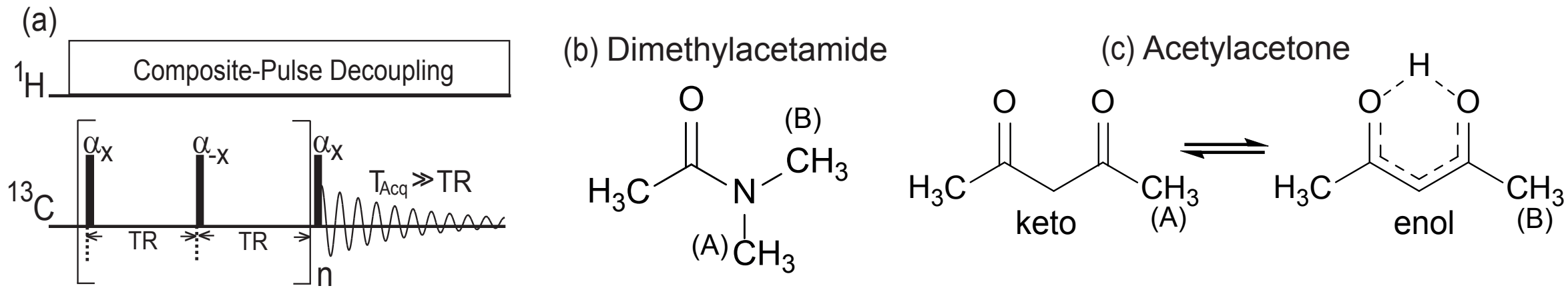

**Figure 7.** (a) Pulse sequence used to monitor the effects of chemical exchange on steady-state NMR experiments. SSFP profiles were acquired with flip angles $\alpha = 10°$ and $\alpha = 23°$. The number of loops $n$ was varied to ensure a constant steady state. Acquisition times $T_{Acq}\sim 1$ s, 4 dummy scans, and WALTZ-16 decoupling[32] (1.8 W power) were used. (b, c) Tautomeric processes followed in this study, involving the interconversion of methyls A and B in DMA and AcAc, as indicated.

## 4. Results

Figure 8 shows SSFP results focusing on methyl interconversions in DMA/$CDCl_3$. These emerge from conventionally-processed proton-decoupled $^{13}C$ NMR acquisitions, whose preparation involved SSFP experiments performed off-resonance (i.e., with the $^{13}C$ carrier frequency placed at an arbitrary position vs the "A" and "B" exchanging methyl peaks). The dotted points in the Figure show the intensities of the exchanging NMR peaks (in magnitude mode) monitored as function of the SSFP's $TR$, for 298, 308 and 318 K. These experimental profiles resemble the profiles of Figure 3, and were fitted to extract *a priori* unknown exchange, chemical shift and relaxation parameters based on the model described above. These fits were done concurrently on both exchanging sites at each temperature after normalizing to a global maximum; further details about the home-developed Matlab® fitting program used to carry out such dynamic fits are discussed in the SI (Section 4). The results of these best-fit SSFP profiles are overlaid as continuous curves in Figure 8, and the emerging best-fit parameters are tabulated in Table I together with their RMSD errors. Further results showing the simpler curves obtained from on-resonance experiments together with their best fits on the same DMA system, are shown in Section 5 of the SI. The ensuing exchange parameters are in good agreement with literature values;[29,30] the emerging offsets of the peaks vs the transmitter also match those derived from the 1D NMR spectrum.

**Table I**: Best-fit exchange parameters for DMA arising from fitting variable-*TR* SSFP data. Experimentally measured chemical shifts of each site are also given for comparison.

| $Temp.$ [K] | $p_A$ | $p_B$ | $k_{ex}$ [$s^{-1}$] | $R_2$ [$s^{-1}$] | $\nu_A$ [Hz] | $\nu_B$ [Hz] | $\nu_A^{exp}$ [Hz] | $\nu_B^{exp}$ [Hz] |
|---|---|---|---|---|---|---|---|---|
| 298 | 0.52 $\pm 0.03$ | 0.48 $\pm 0.03$ | 1.0 $\pm 0.24$ | 1.7 $\pm 0.13$ | 145.9 $\pm 0.1$ | $-319$ $\pm 0.3$ | 146.2 | -318.8 |
| 308 | 0.51 $\pm 0.03$ | 0.49 $\pm 0.03$ | 2.7 $\pm 0.26$ | 2.6 $\pm 0.18$ | 143.7 $\pm 0.2$ | $-319.2$ $\pm 0.2$ | 144.5 | -318.6 |
| 318 | 0.51 $\pm 0.03$ | 0.49 $\pm 0.03$ | 6.7 $\pm 0.67$ | 4.1 $\pm 0.39$ | 137.7 $\pm 0.2$ | $-323.9$ $\pm 0.2$ | 139.1 | -322.7 |

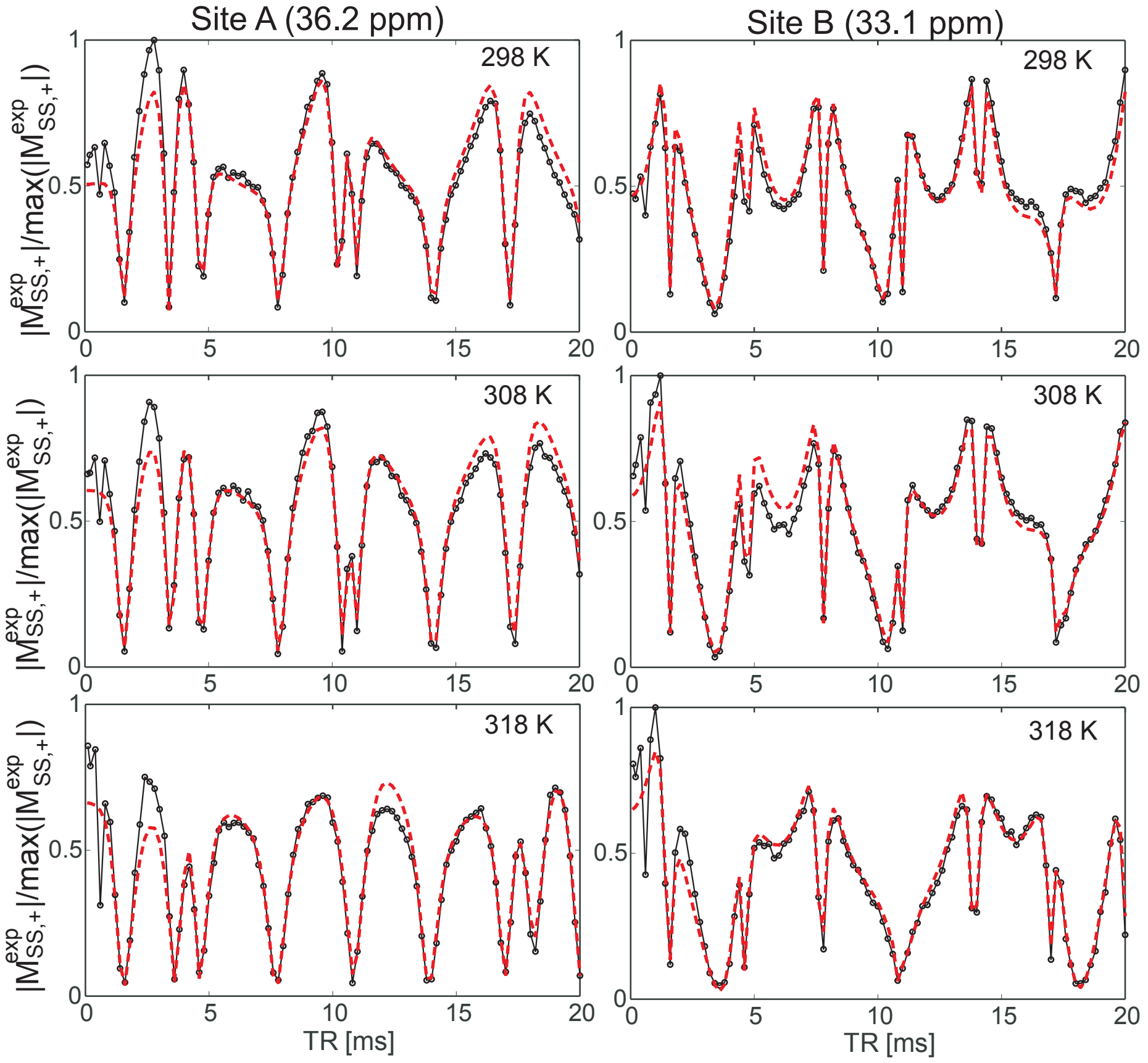

**Figure 8.** SSFP profiles arising from DMA's exchanging $^{13}C$ sites (labelled A and B) recorded at various temperatures with a 10° flip angle. TR was varied from 0.1 to 20.1 ms in 0.2 ms steps. Black dots denote experimental results and red dashed lines are best simultaneous fits to the site A and site B data sets for each temperature, based on the parameters listed in Table I. The small dips deviating from the fits at 140 µs (4th point from the start) arise from an interference between the SSFP and the WALTZ-16 decoupling sidebands[27] and hence were not captured by the fits.

A second set of experimental and best-fit SSFP profiles, this time targeting the keto-enol tautomerism of acetylacetone (AcAc) in DMSO-d6/NaOH at two temperatures, is displayed in Figure 9. The fit focused jointly on the $^{13}$C SSFP magnetizations of the keto (30.3 ppm) and enol (24.2 ppm) methyl groups as a function of $TR$. 50 experiments were performed starting from $TR = 100\ \mu s$ to 10.1 ms in equal steps –this time with a sensitivity-optimized flip angle of $23^o$. A noticeable difference between these SSFP profiles and those in Figure 8 is their sharp drop at low TRs, which on the basis of simulations (Fig. 4) is expected to happen as $k_{ex}$ is in this case larger. Also the larger flip angles lead to such drops of the SSFP signal at short $TR$s (see SI Section 6 for an example). The best-fit set of parameters arising from fitting these off-resonance SSFP experiments is given in Table II. Once again there is good agreement with independent literature and spectral data;[30] further experiments performed on resonance at one of the sites, and at different temperatures along with their best fit parameters, are presented in the SI (Section 6).

**Table II**: Best-fit exchange parameters for AcAc's methyl groups. Experimentally measured chemical shifts of each site are also given for comparison.

| $Temp.$ [K] | $p_A$ | $p_B$ | $k_{ex}$ [s$^{-1}$] | $R_2$ [s$^{-1}$] | $\nu_A$ [Hz] | $\nu_B$ [Hz] | $\nu_A^{exp}$ [Hz] | $\nu_B^{exp}$ [Hz] |
|---|---|---|---|---|---|---|---|---|
| 293 | 0.27 ±0.07 | 0.73 ±0.19 | 12.1 ±1.39 | 4.3 ±0.7 | 277.9 ±0.4 | −648.9 ±0.4 | 281.9 | −646.6 |
| 310 | 0.29 ±0.06 | 0.71 ±0.16 | 48.4 ±2.7 | 3.2 ±0.27 | 257.1 ±0.8 | −665.7 ±0.5 | 259.4 | −665.2 |

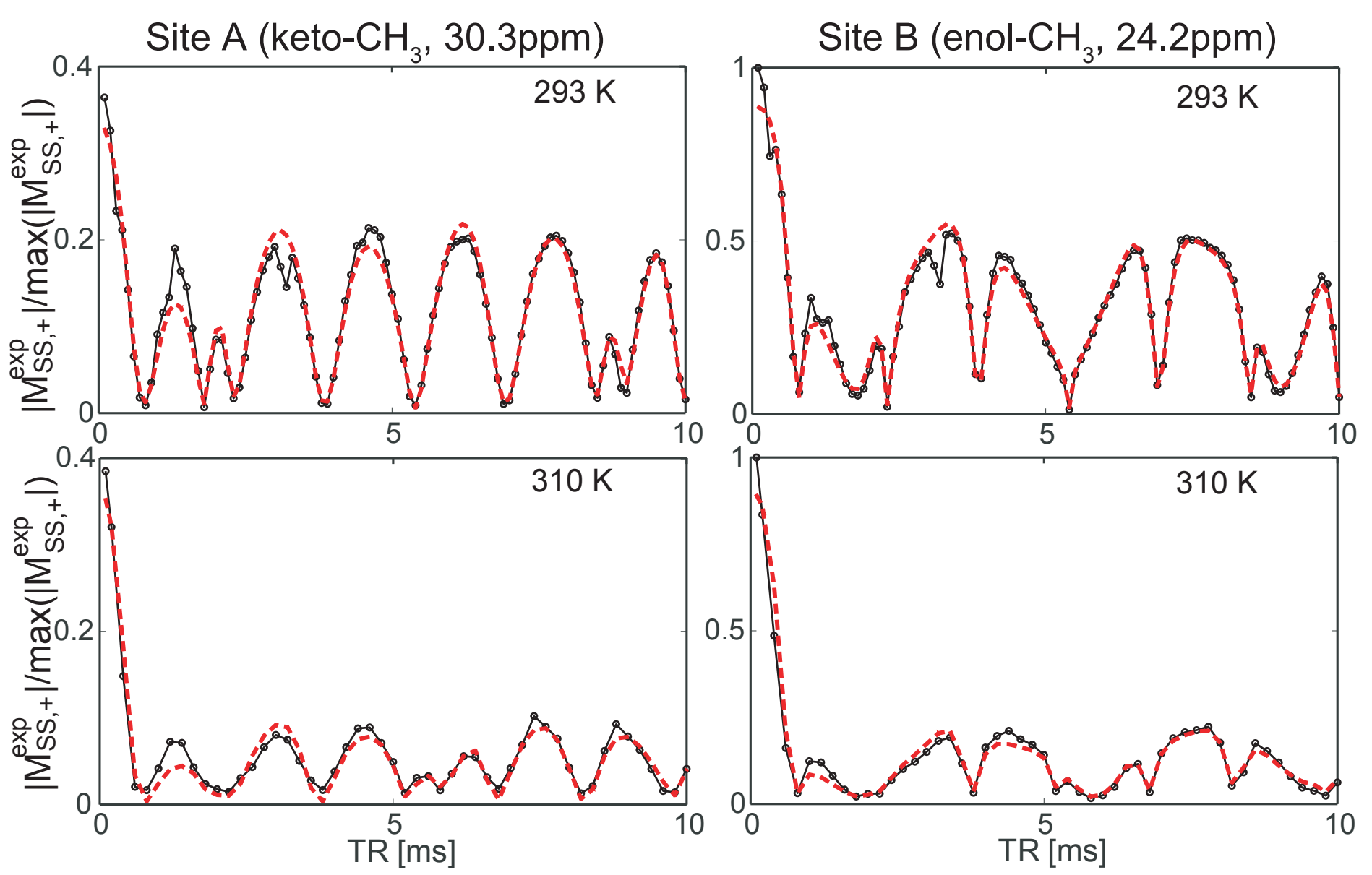

**Figure 9.** Experimental (dots) and best-fit (red) SSFP profiles for AcAc's methyl peaks at two temperatures, measured as a function of a $TR$ varied from 0.1 to 10.1 ms in 0.2 ms steps with an $\alpha = 23°$ flip angle. The offsets of sites A and B with respect to carrier frequency are indicated in Table II, as are other best-fit parameters. The minor dip observed at $TR = 3.2$ ms for both sites at 293 K originates once again from an interference with the WALTZ-16 proton decoupling.

It is interesting to assess the errors of these measurements, particularly as the exchange rates and $R_2$ values arising for fitting measurements carried out at different offsets are slightly different –while chemical shifts and populations are nearly identical (cf. Tables I and II with SI, Section 6). The manner in which errors were assessed was on the basis of $\chi^2$ plots, which are illustrated in Figures 10a and 10b for DMA (298 K) and AcAc (293 K) respectively (see SI, Sections 6 and 7 for similar $\chi^2$ plots at other temperatures and different offsets). In general these plots show sharp, well-defined values for shifts and populations, but relatively shallow minima with correlated $R_2$ and $k_{ex}$ values for all fits.  This is to be expected based on insight from 1D line shape experiments, where in such slow-exchange regimes excess line widths from transverse relaxation and from exchange broadening are hard to discern from one another. Indeed, when comparing the kinetics emerging from data taken at identical temperatures but different offsets for either of the two compounds, the fits tend to reduce the effects of $R_2$ relaxation at the expense of slight increases in the exchange rate values. At this point it is not clear if these changes might reflect imperfections arising from the RF pulses –foremost RF inhomogeneities affecting differently the apparent $R_2$s under on- or off-resonance conditions– or biases in the assumptions taken to perform the SSFP fits (e.g., identical $R_1$s and $R_2$s for all exchanging sites). Further examination of these features is in progress.

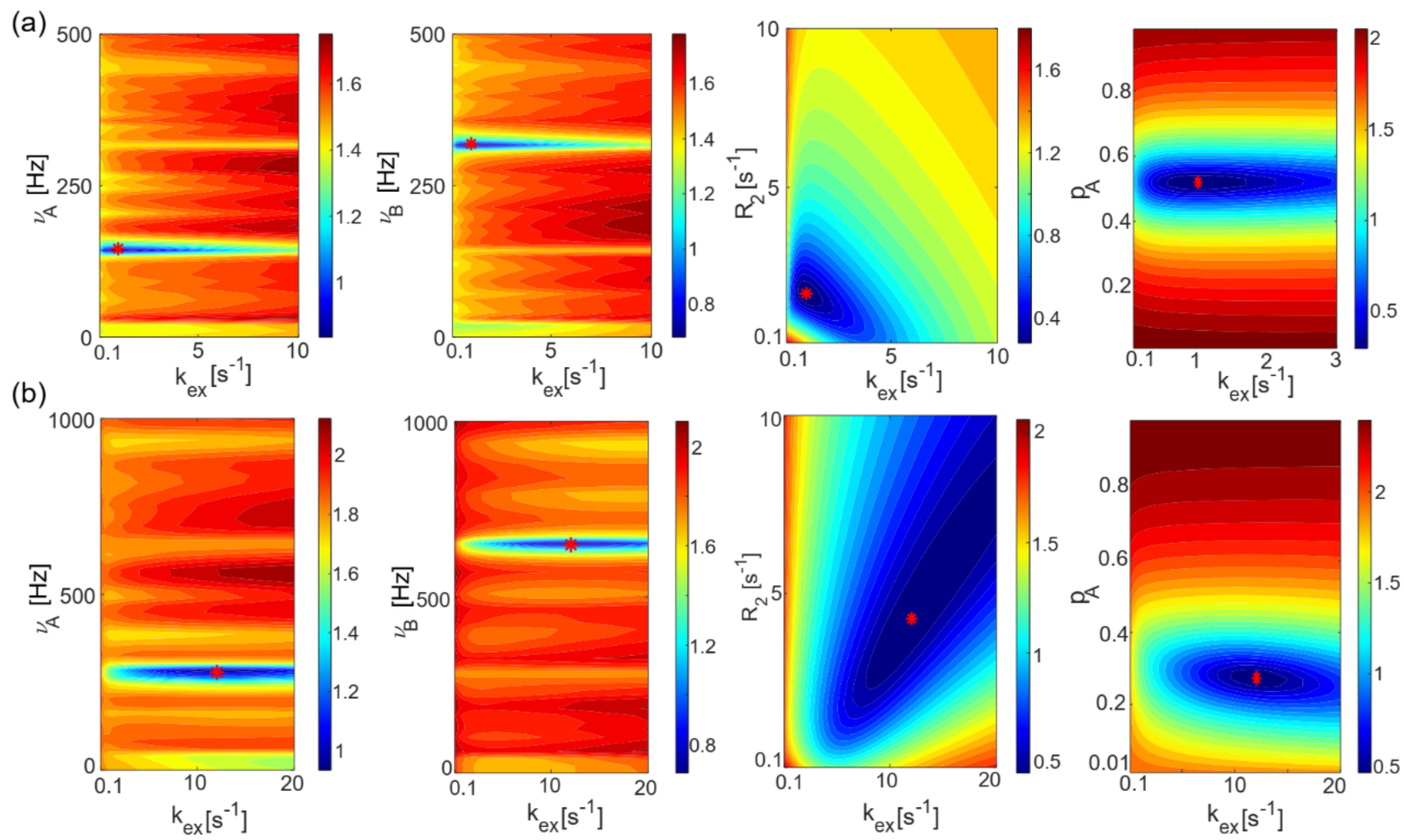

**Figure 10.** $\chi^2$ hot-cold maps showing the correlations between errors extracted for DMA (a) and AcAc (b), when focusing on the $k_{ex}$ vs $\nu_A$, $k_{ex}$ vs $\nu_B$, $k_{ex}$ vs $R_2$, and $k_{ex}$ vs $p_A$ planes. The boundaries considered for the search leading to these best-fits were: (a) $p_A = 0.01$ to 0.99, $k_{ex} = 0.1$ to 10 $s^{-1}$; $R_2 = 0.1$ to 10 $s^{-1}$, $\nu_A = 0$ to 500 Hz; $\nu_B = 0$ to 500 Hz.  (b) $p_A = 0.01$

to 0.99, $\nu_A = 0$ to 1000 Hz; $\nu_B$ = 0 to 1000 Hz, $k_{ex} = 0.1$ to 20 $s^{-1}$, and $R_2 = 0.1$ to 10 $s^{-1}$. All fits assumed $R_{1A} = R_{1B} = R_1 = 0.1\ s^{-1}$ (fixed) for all sites, and $R_{2A} = R_{2B} = R_2$. The best fit parameters in all plots are marked with a "*".

## 5. Discussion and Conclusions

SSFP is widely used in MRI, where for the case of a single-site resonance like that arising from water it is often a method of choice thanks to its superior sensitivity. The method has also been explored based on $R_{1\rho}$ models as an optimal way to obtain CEST MRI images;[23] however, its outcome under the classical McConnell-Bloch assumptions of chemical exchange between chemically resolved sites, has not yet been fully explored. The present study did a first step in this direction, by developing a model for multiple discrete sites exchanging in the slow and fast regimes. These Liouville-space theoretical derivations lead analytical two-site exchange treatments applicable in the slow- and fast-exchange regimes, which agreed well with "brute force" numerical calculations applicable to arbitrary time scales, and with experimental measurements that focused on two-site dynamics proceeding slow in the NMR timescale. These measurements were purposedly simple, and involved achieving steady states as a function of *TR* without real-time acquisition of the emitted signals, followed by a conventional 1D $\{^1H\}^{13}C$ NMR observation. This provided data with an unencumbered, constant high resolution, while allowing us to use *TR* as a kinetic extraction parameter –akin to the $\tau_{CPMG}$ used in RD experiments. The kinetic information will also be present in directly detected SSFP acquisitions –i.e., in the signal sampled in-between the pulses– as well as in the recently proposed phase-incremented SSFP proposals for high resolution NMR.[17,18] Ongoing investigations will present routes to access kinetic parameters while relying on these faster, more efficient acquisition strategies.

Several additional simplifications characterized this study, which are also worth highlighting. These involved the aforementioned disregard of potential differences in the $R_1/R_2$ relaxation rates between the exchanging sites, and neglecting the potential effects of RF inhomogeneity. Also ignored was the potential usefulness of the flip angle α as independent variable, complementing the variable-*TR* information for a more accurate extraction of the kinetic parameters. The present study also focused on arguably the simplest exchanging cases available in the NMR dynamics literature, involving two-site tautomerisms in DMA and AcAc taking place on the slow exchange regime. Simulations suggest that the SSFP approach could tackle much more complex cases of exchange in a much wider variety of rate scenarios –once again by bringing together beneficial aspects known from CEST and CPMG dynamic experiments. Another simplification taken by this study involved its focus on $^{1}$H-decoupled $^{13}$C NMR at natural abundance. These isolated spin-1/2 systems helped us bypass additional sources of SSFP profile changes that we have observed and which may arise if focusing on more abundant spins like $^{1}$Hs, where complex shapes vs *TR* may also derive from NOE and/or homonuclear J-coupling effects. While potentially interfering with the kinetic aspects that we sought to highlight in this work, these NOE and J-driven effects carry valuable information of their own, whose potential extraction via SSFP also remains to be fully exploited. These and other hitherto rich effects arising in SSFP NMR, will be the focus of future investigations.

**Acknowledgments:** Support from the Israel Science Foundation (grant 1874/22), the Minerva Foundation, the ERC Advanced Grant Project 101200719 "SteadyNMR", and the Perlman Family Foundation, are acknowledged. SJ acknowledges Weizmann Institute for a sabbatical fellowship; ZO thanks the Azrieli Foundation for a postdoctoral fellowship.

## References

1. D. M. Grant and R. K. Harris Eds., Encyclopedia of Nuclear Magnetic Resonance, Vols. 1–10 (John Wiley & Sons, Ltd., Chichester, United Kingdom, 2012).
2. N. Bloembergen, E. M. Purcell, R. V. Pound, Relaxation effects in nuclear magnetic resonance absorption. Physical Review. **73**, 679-712 (1948).
3. L. M. Jackman, F. A. Cotton, Eds., Dynamic NMR Spectroscopy, Academic Press, New York, (1975).
4. I. J. Kaplan, G. Fraenkel, NMR of Chemically Exchanging Systems, Academic Press, New York, (1980).
5. A. D. Bain, Chemical exchange in NMR. Progress in Nuclear Magnetic Resonance Spectroscopy **43**, 63-103 (2003)
6. H. M. McConnell, Reaction Rates by Nuclear Magnetic Resonance. J. Chem. Phys. **28**, 430 (1958).
7. I. J. Kaplan, Exchange Broadening in Nuclear Magnetic Resonance, J. Chem. Phys. **28**(2), 278–282 (1958).
8. S. Alexander, Exchange of Interacting Nuclear Spins in Nuclear Magnetic Resonance. I. Intramolecular Exchange, J. Chem. Phys. **37**(5), 967–974 (1962).
9. Z. Luz, and S. Meiboom, Nuclear Magnetic Resonance Study of the Protolysis of Trimethylammonium Ion in Aqueous Solution—Order of the Reaction with Respect to Solvent. J. Chem. Phys., **39** (2), 366–370 (1963).
10. S. Forsen, and R. A. Hoffman, Study of Moderately Rapid Chemical Exchange Reactions by Means of Nuclear Magnetic Double Resonance, J. Chem. Phys., **39**(11), 2892-2901 (1963).
11. K. M. Ward, and R. S. Balaban, Determination of magnetic resonance parameters of proton exchange groups, the basis for CEST. Magn. Reson. Med. **44**, 799–802 (2000).
12. N. L. Fawzi, J. Ying, D. A. Torchia, G. M. Clore, Probing exchange between massive complexes and visible states by DEST. J. Am. Chem. Soc. **133**, 9964–9967 (2011).

13. P. Vallurupalli, G. Bouvignies, and L. E. Kay, Measurement of invisible excited states of proteins. Nature Protocols **7**, 1913-1935 (2012).
14. A. Allerhand, and H. S. Gutowsky, Spin-Echo NMR Studies of Chemical Exchange. J. Chem. Phys. **41**, 2115-2126 (1964).
15. V. Z. Miloushev, and A. G. Palmer, III, R(1rho) relaxation for two-site chemical exchange: general approximations and some exact solutions. Journal of Magnetic Resonance **177**, 221-227 (2005)
16. A. G. Palmer, A. G. Chemical exchange in biomolecules: relaxation dispersion and saturation transfer. Annu. Rev. Biophys. **43**, 129–149 (2014).
17. T. He, Y. Zur, E. T. Montrazi, and L. Frydman, Phase-Incremented Steady-State Free Precession as an Alternate Route to High-Resolution NMR, J. Am. Chem. Soc. **146**, 3615-3621 (2024).
18. M. Shif, Y. Zur, A. Lupulescu, T. He, E. T. Montrazi, and L. Frydman Maximizing spectral sensitivity without compromising resolution in phase-incremented, steady-state solution NMR, Nat. Commun. **16**, 5745 (2025).
19. H. Y. Carr, Steady-State Free Precession in Nuclear Magnetic Resonance. Phys. Rev. **12** (5), 1693–1700 (1958).
20. K. Scheffler, and S. Lehnhardt, Principles and applications of balanced SSFP techniques. Eur. Radiol. **13**, 2409–2418 (2003).
21. O. Bieri, and K. Scheffler, Fundamentals of balanced steady state free precession MRI. J. Magn. Reson. Imaging **38**, 2–11 (2013).
22. T. Yuwen, L. E. Kay, and G. Bouvignies, Dramatic Decrease in CEST Measurement Times Using Multi-Site Excitation. ChemPhysChem., **19** (14), 1707–1710 (2018).
23. S. Zhang, Z. Liu, A. Grant, J. Keupp, R. E. Lenkinski, and E. Vinogradov, Balanced steady-state free precession (bSSFP) from an effective field perspective: Application to the detection of chemical exchange. J. Magn. Reson. **275**, 55-67 (2017).
24. Y. Zur, S. Stokar, and P. Bendel, An Analysis of Fast Imaging Sequences with Steady-State Transverse Magnetization Refocusing. Magn. Reson. Med. **6**, 175–193 (1988).
25. J. Cavanagh, J. F. Wayne, M. Rance, N. J. Skelton, and A. G. Palmer III, Protein NMR Spectroscopy: Principles and Practice, 2nd ed. (Academic Press, Elsevier, Amsterdam, 2007).
26. Y. Jin, Y .Cui, T. Yuwen, NMR methods for investigating functionally relevant biomolecular dynamics, Magn. Reson. Lett., **5**, 2025, 200195
27. S. Jayanthi, Z. Osifová, M. Shif, A. Lupulescu, and L. Frydman, Steady-state free precession NMR in the presence of heteronuclear couplings and decoupling: More than meets the eye, J. Chem. Phys., **163**, 154201 (2025).
28. L. W. Reeves, R. C. Shaddick, and K. N. Shaw, Nuclear Magnetic Resonance Studies of Multi-site Chemical Exchange III., Hindered rotation in dimethylacetamide, dimethyl trifluoroacetamide, and dimethyl benzamide, Can. J. Chem., **49**, 3683-3691 (1971)
29. F. P. Gasparro, N. H. Kolodny, NMR Determination of the rotational barrier in N,N-dimethylacetamide, A physical chemistry experiment, J. Chem. Edn., **54**(4), 258-261 (1977).
30. E. J. Drexler, K. W. Field, An NMR study of keto-enol tautomerism in β-dicarbonyl compounds, J. Chem. Edn.,53(6), 392-393 (1976).
31. L. W. Reeves, Nuclear magnetic resonance measurements in solutions of acetylacetone. The effect of solvent interactions on the tautomeric equilibrium, Can. J. of Chem., 35, 1351-1365 (1957)

32. A. J. Shaka, J. Keeler, R. Freeman, Evaluation of a new broadband decoupling sequence: WALTZ-16, J. Magn. Reson., **53**(2) , 313-340 (1983).

**Supporting Information for**
**Unravelling Chemical Exchanges Through Steady State Free Precession NMR**

Sundaresan Jayanthi, Adonis Lupulescu, Mark Shif, Zuzana Osifová, and Lucio Frydman*

Department of Chemical and Biological Physics, Weizmann Institute of Science, Rehovot, Israel

## Section - 1: SSFP and slow exchange – Analytical magnetization expressions for an on-resonance site

In this calculation we consider a two-site exchange system between sites A and B, assumed to have the same longitudinal and transverse relaxation rates: $R_{1A} = R_{1B}, R_{2A} = R_{2B}$. We found that for site A on-resonance, the steady-state transverse magnetization component $M_X(A) = 0$ and $M_Y(A)$ are

$$M_Y(A) = \frac{2p_A(E_1 - 1)\, sin(\alpha)\, \Sigma}{\Theta_1 + \Theta_2 \cos(2\alpha) + \Theta_3 \cos^2(\frac{\alpha}{2})\, cos(\omega_B TR) + 2cos\alpha\, [\Theta_4 - \Theta_5\, 2E_2 \cos^2\left(\frac{\alpha}{2}\right) \cos(\omega_B TR)]} \quad (S1.1)$$

where the different constants $\Sigma, \Theta_1, \Theta_2, \Theta_3, \Theta_4, \Theta_5$ are,

$$\Sigma = e^{k_{ex}TR} e^{2k_{ex}p_B TR} - E_1 E_2^2 - E_1 e^{2k_{ex}p_B TR} cos(\alpha) + E_2^2 e^{k_{ex}TR} cos(\alpha) - E_2 e^{k_{ex}p_B TR}(e^{k_{ex}TR} - E_1)(1 + cos(\alpha))cos(\omega_B TR).$$

$$\Theta_1 = (2e^{k_{ex}TR} + E_1^2)\, e^{2k_{ex}p_B TR} + E_2^3\, (e^{k_{ex}p_B TR} + 2E_1^2 e^{-k_{ex}p_A TR}) + E_1 E_2(-2 + p_A)(e^{-k_{ex}p_A TR} e^{2k_{ex}p_B TR} + E_2 e^{k_{ex}TR}) - E_1 E_2(1 + p_A)\, (e^{3k_{ex}p_B TR} + E_2).$$

$$\Theta_2 = E_1^2\, e^{2k_{ex}p_B TR} + E_2^3 e^{k_{ex}p_B TR} - p_B E_1 E_2(E_2 + e^{3k_{ex}p_B TR}) - E_1 E_2 p_A(e^{-k_{ex}p_A TR} e^{2k_{ex}p_B TR} + E_2 e^{k_{ex}TR}).$$

$$\Theta_3 = 4e^{-k_{ex}p_A TR}\, E_2\Big(e^{2k_{ex}TR}(1 - E_1 p_B) - e^{k_{ex}TR} E_1 p_A(1 + E_2 e^{k_{ex}p_B TR} + e^{k_{ex}p_B TR} E_1 E_2(E_1 - p_B)\Big).$$

$$\Theta_4 = (-e^{2k_{ex}p_B TR} E_1 - e^{k_{ex}TR} e^{2k_{ex}p_B TR} E_1 + e^{3k_{ex}p_B TR} E_2 + e^{-k_{ex}p_A TR} e^{2k_{ex}p_B TR} E_1^2 E_2 + e^{k_{ex}TR} E_2^2 + E_1^2 E_2^2 - e^{-k_{ex}p_A TR} E_1 E_2^3 - e^{k_{ex}p_B TR} E_1 E_2^3).$$

$$\Theta_5 = (e^{2k_{ex}p_B TR} E_2(1 - E_1 p_B) - e^{k_{ex}TR} e^{k_{ex}p_B TR} E_1 p_A - e^{-k_{ex}p_A TR} e^{k_{ex}p_B TR} E_1 E_2 p_A + e^{k_{ex}p_B TR} E_1(E_1 - p_B)).$$

These equations are lengthy and do not provide much physical insight; however, they provide an efficient starting point for fitting programs that analyze experimental SSFP profiles to arrive at plausible exchange parameters. The above equation is specific to the slow exchange approximation, and the main text shows how it deviates from the actual behavior at small TR's.

Eq. (S.1.1) becomes more complex if we include the situation where $R_{1A} \neq R_{1B}$, and $R_{2A} \neq R_{2B}$. A glimpse of the complexity of the modified equation is shown below:

$$M_Y(A) = \left(p_A e^{(-TR(G-k_{ex})/2)} sin\alpha \left(2E_{1B}^2 GXe^{(TR(G+2X+Z+3kex+4p_A k_{ex})/2)} \ldots\ldots\right)\right) \\ /2E_{1B}e^{(TR(Z-G+2k_{ex})/2)} \ldots\ldots)) \quad (S1.2)$$

where, $E_{1A} = e^{-R_{1A}TR}$, $E_{2A} = e^{-R_{2A}TR}$, $E_{1B} = e^{-R_{1B}TR}$, $E_{2B} = e^{-R_{2B}TR}$

$$X = \sqrt{(k_{ex} + (R_{1A} - R_{1B}))^2 - 4p_A k_{ex}(R_{1A} - R_{1B})}$$

$G = X + k_{ex} + R_{1A} - R_{1B}$, $Z = X - k_{ex} + R_{1A} - R_{1B} \ldots$

The overall forms of these equations as provided by Matlab®, are too lengthy to be here included but can be supplied upon request.

## Section 2: SSFP in the presence of slow exchange: Two sites with arbitrary offsets.

Symbolic tools also allowed us to derive analytical equations for arbitrary offsets of the two exchanging sites even when $R_{1A} \neq R_{1B}$ and $R_{2A} \neq R_{2B}$, for the $k_{ex} \ll |\omega_A - \omega_B|$ situation. These are again too lengthy to be conveniently displayed; here we display a few terms to show the main factors involved (further material available upon request).

$$M_X(A) = E_{2A}e^{TRk_{ex}(p_A-1)} sin(\omega_A TR) sin\alpha(4QGp_A^2 e^{2TRp_A k_{ex}} cos\alpha - 2PXYe^{2p_A TRk_{ex}} \\ + \cdots/\ldots.)\ etc. \quad (S2.1)$$

$$Y = \sqrt{E_{1A}E_{1B}} e^{-TR(X+k_{ex})/2},$$

$$P = p_A(X - k_{ex} + R_{1A} - R_{1B})(1 + e^{XTR}) - 2p_A X/Y$$

$$Q = e^{-TR(X+k_{ex}+R_{1A}+R_{1B})/2}\left(X - k_{ex} - R_{1A} + R_{1B} + e^{XTR}(X + k_{ex} + R_{1A} + R_{1B}) \\ - 2Xe^{TR(R_{1A}+R_{1B}+X+k_{ex})/2}\right)$$

$$M_Y(A) = sin\alpha(E_{2A}e^{TRk_{ex}(p_A-1)} cos(\omega_A TR) \\ - 1)(4GQp_A^2 e^{2TRp_A k_{ex}} cos\alpha)(4QGp_A^2 e^{2TRp_A k_{ex}} cos\alpha - 2PXYe^{2p_A TRk_{ex}} \\ + \cdots/\ldots.)\ etc. \quad (S2.2)$$

$$M_X(B) = E_{2B}e^{TRp_A k_{ex}} sin(\omega_B TR) sin\alpha/(e^{2TRp_A k_{ex}} cos\alpha + E_{2B}^2 cos^2(\omega_B TR) sin^2\alpha + \ldots) \\ /\ldots..\ etc. \quad (S2.3)$$

$$M_Y(B) = -e^{TRp_A k_{ex}} sin\alpha(e^{TRp_A k_{ex}} - E_{2B} cos(\omega_B TR)) \\ /\left(e^{2TRp_A k_{ex}} cos\alpha(e^{2TRp_A k_{ex}} cos\alpha + E_{2B}^2 cos^2(\omega_B TR) sin^2\alpha + \ldots)\right) \\ /\ldots..\ etc. \quad (S2.4)$$

## Section 3: Validity of the analytical SSFP derivations against numerical simulations at short TR's: Slow, intermediate and fast exchanges.

Figure S1a shows "brute force" against analytical predictions for the SSFP signal as function of TR, as the slow exchange evolves into the intermediate regime. Notice the deviations arising upon considering short $TR$s, for which the neglection of off-diagonal $k_{AB}$ and $k_{BA}$ matrix elements vs evolution in the transverse part of $\mathcal{L}$ stops being justified. Figure S1b complements this with an extension (for identical shift and relaxation parameters but a different carrier offset) of the SSFP curves from the intermediate to the fast-exchange regime. Once again, deviations between analytical and "brute force" SSFP profiles occur at low TR values, and diminish with increasing $k_{ex}/\Delta\omega_{Ab}$.

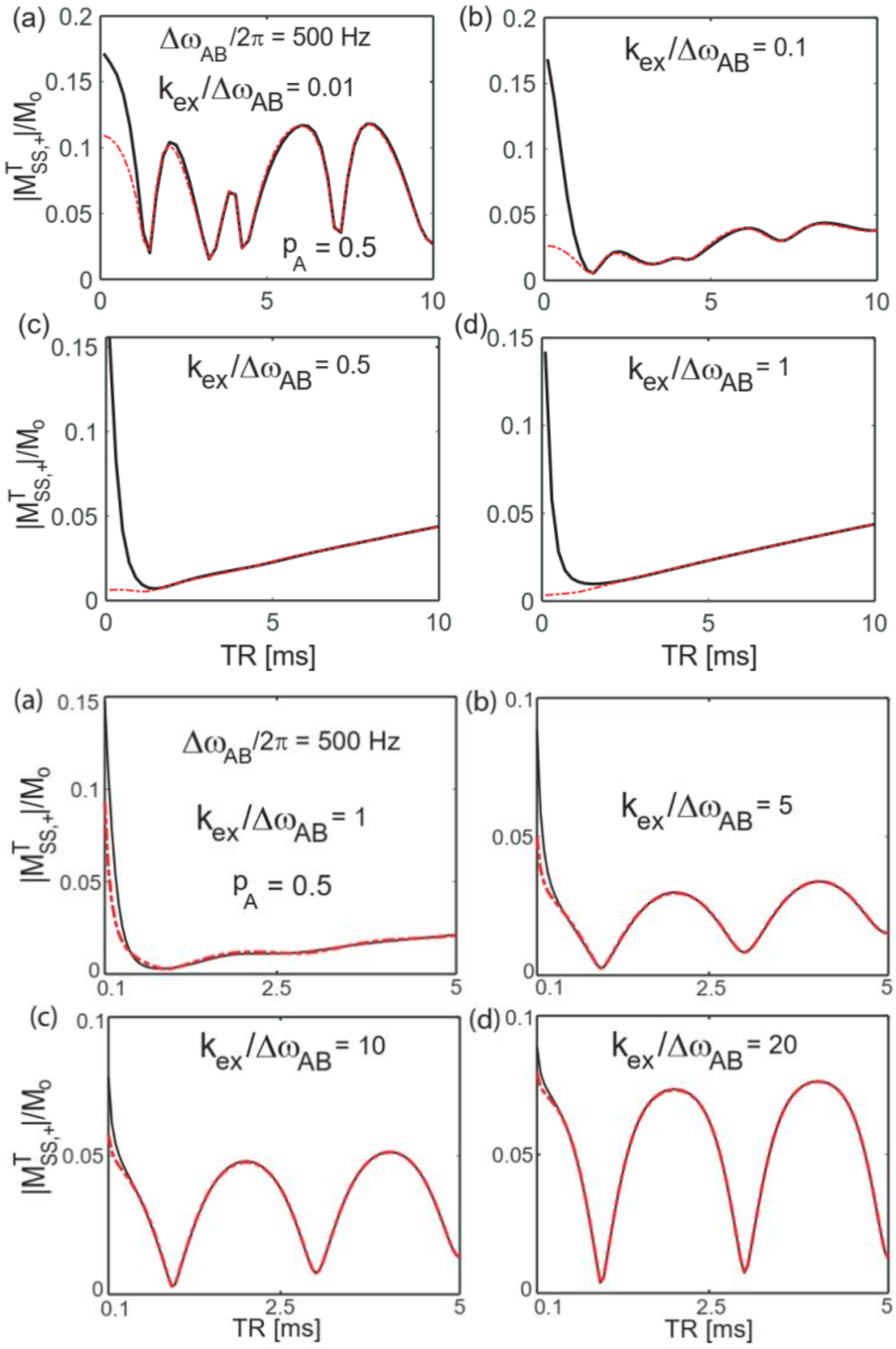

**Figure S1a**: Total magnetization SSFP profiles as function of $k_{ex}/\Delta\omega_{AB}$, in the slow exchange regime. Numerical (black) and analytical (red) profiles are confronted. The offsets of sites A and B vs the carrier are 150 and -350 Hz, respectively. Parameters used in the simulation are $\alpha = 20^o, R_{1A} = R_{1B} = 1\, s^{-1}, R_{2A} = R_{2B} = 5\, s^{-1}$.

**Figure S1b**: Total magnetization SSFP profiles as function of $k_{ex}/\Delta\omega_{AB}$, in the fast exchange regime. Numerical (black) and analytical (red) profiles are confronted. The offset of the (now exchange-averaged) observed peak vs the carrier, is here 500 Hz. Other parameters used are $\alpha = 20^o, R_{1A} = R_{1B} = 1\, s^{-1}, R_{2A} = R_{2B} = 5\, s^{-1}$.

## Section-4: SSFP in the fast chemical exchange regime: Rates, populations, and flip angles.

The dependence of SSFP profile and its corresponding derivative for two flip angles, as a function of population for a given exchange rate, for various exchange rates, and finally as a function of flip angles for two different populations are shown. In all cases, derivatives of the SSFP profiles provide better sensitivity with respect to the varied parameter. Remarkably, SSFP profiles and their derivatives at fast exchange rates are more sensitive to skewed populations.

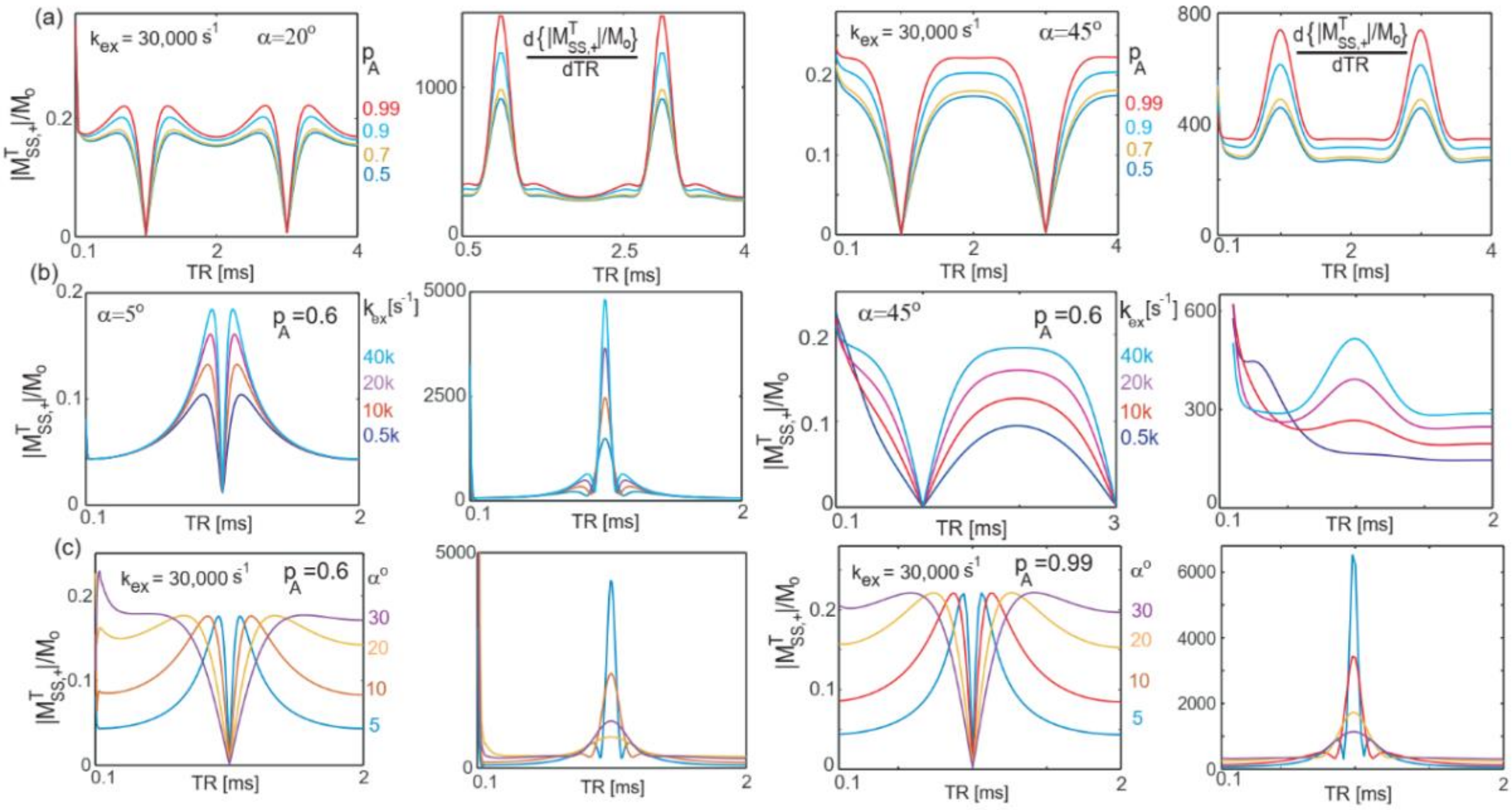

**Figure S2**: SSFP profiles and their derivatives as a function of (a) population for flip angles α = $20^o$ and $45^o$ (b) as a function of exchange rates for $\alpha = 5^o$ and $45^o$, (c) as a function of flip angles for two different populations of site A, 0.6 and 0.99. The assumed relaxation rates were $R_{1A} = R_{1B} = 1\ s^{-1}, R_{2A} = R_{2B} = 5\ s^{-1}$.

## Section 5: Details of the "SSFP Dynamics" fitting program: How to use it and what it yields

**5.1 The input data.** As part of this study we make available a custom-made "SSFP Dynamics" fitting software that we wrote to analyze up to three-site exchanges from variable-TR 1D Bruker NMR data. This is available at https://github.com/MarkS098/SSFP-dynamics. As input to such processing one needs a set of one-dimensional NMR data collected as a function of TR; this can be directly 1D spectra, or FIDs that were processed from Bruker formatted datasets using a custom MATLAB pipeline (raw_data_process.m). For each experiment read as time-domain FIDs a truncation of the initial group delay points was performed (not needed if inputting pre-processed frequency-domain real/imaginary or magnitude mode data). Spectral intensities were then analyzed from the ensuing variable-TR set within predefined chemical shift windows corresponding to the regions of interest. Peak amplitudes were extracted using a local maximum search constrained by a user defined minimum intensity threshold to ensure robustness against noise. The repetition time ($TR$) associated with each experiment was also extracted from the acquisition parameters provided by Bruker's standard experimental data format; also the associated repetition times and flip angles of the experiments were read from the data sets

**5.2 How the fitting works** The fitting part of the software package ("ssfp_exchange_joint_fit.m") starts by having the user input the data, together with an arbitrary number of points to omit from the variable-TR signal fitting (e.g., the initial one if corrupted). If measured or estimated, the common longitudinal recovery rate $R_1$ to all sites needs also to be chosen. Data generated for the exchanging sites from the processing pipeline is loaded into the fitting program containing the intensity and $TR$ information for the exchanging peaks, then parameter boundaries are set for the following parameter vector by the user:

$$\theta = [M_{A0}, M_{B0}, M_{C0}, k_{AB}, k_{BC}, k_{AC}, \nu_A, \nu_B, \nu_C, R_2] \quad \text{(S5.1)}$$

where $M_{i0}$represent the fractional populations such that $\sum M_{i0} = 1$, $k_{i,j}$ are the exchange rate constants, $\nu_i$ are the chemical shift frequency offsets and $R_2$ is the (also common) transverse recovery rate. All these parameters will be simultaneously fitted from the experimental data. To this end, the SSFP signal is modelled using a generalized two or three site chemical exchange simulator ("chem_exchange_sim.m") based on the Bloch-McConell equations, which govern the evolution of the magnetization vector $\mathrm{M}_{exp}$ in a multi-site exchange system. The simulated signal $\mathrm{M}_{sim}$, is defined as a function of $TR$ and of a parameter vector $\theta$ to be found.

**5.3 Formulating the fit and optimizing its minimum.** The search for the parameter vector $\theta$ is treated as a bounded, nonlinear least squares optimization problem. Population normalization is enforced such that $M_{A0} + M_{B0} + M_{C0} = 1$ with an automatic reduction to a two-site model when site C is inactive. The simulated signals from all observable sites $M_{sim,A}, M_{sim,B}$ ... are concatenated into a single global vector model $\mathrm{M}_{sim}$, and evaluated against a similarly concatenated experimental data vector $\mathrm{M}_{exp}$. To account for arbitrary global scaling differences between the experimental signal, $\mathrm{M}_{exp}$, and the simulated signal, $\mathrm{M}_{sim}(\theta)$, an analytical scale factor $c(\theta)$ is computed at each objective function evaluation via an orthogonal projection:

$$c(\theta) = \frac{\mathrm{M}_{sim}(\theta)^T \, \mathrm{M}_{exp}}{\mathrm{M}_{sim}(\theta)^T \, \mathrm{M}_{sim}(\theta)} \qquad \text{(S5.2)}$$

The scaled residual vector is then defined to penalize discrepancies between the model and the data:

$$r(\theta) = \frac{c(\theta)\mathrm{M}_{\mathrm{sim}}(\theta) - \mathrm{M}_{\mathrm{exp}}}{\frac{1}{N}\sum_{i=1}^{N} \mathrm{M}_{\mathrm{exp,i}}} \qquad \text{(S5.3)}$$

where $N$ is the number of acquired $TR$ data points. Normalization by the mean of the signal amplitude is introduced to make the residuals scale-invariant across datasets. The optimization problem seeks the optimal parameter set $\hat{\theta}$ that minimizes the $L_2$ norm of the residuals subject to lower (lb) and upper (ub) bounds input by the user:

$$\hat{\theta} = \arg \min_{\mathrm{lb} \leq \theta \leq \mathrm{ub}} \frac{1}{2} \left|\left|r(\theta)\right|\right|_2^2 \qquad \text{(S5.4)}$$

To guarantee convergence to a global minimum, the fitting program utilizes MATLABs MultiStart function combined with Latin Hypercube Sampling (LHS). Unlike pure Monte Carlo initialization, which can leave regions of the parameter space unexplored due to random clustering, LHS stratifies each parameter dimension into equally probable intervals, ensuring a highly uniform, space filling distribution of $N_{start}$ initial guesses across the bounded multidimensional parameter space. From each of these $N_{start}$ uniformly distributed initial guesses, a local Trust-Region-Reflective solver ("lsqnonlin") is deployed. By looking at the optimization problem from multiple different starting points, the algorithm reliably identifies the global minimum $\hat{\theta}$ even in the presence of multiple similar minima or partial parameter degeneracy. This is important because in certain cases the $\chi^2$ error surface for multi-site SSFP exchange can be highly non-convex. In particular, the correlation between the $R_2$ and $k_{ex}$ frequently creates long flat "valleys" in the optimization landscape, trapping standard gradient descent algorithms in local minima.

**5.4 Error Estimation via Non-Parametric Bootstrap and Error Surface Mapping.** Standard error estimation relies on the inversion of the Jacobian-derived Fisher Information Matrix ($I = J^T J$). To rigorously quantify the error in the fitted parameters –in particular with respect to $R_2$ and $k_{ex}$, which can become collinear resulting in a near singular Jacobian– a non-parametric bootstrap protocol is utilized. This method evaluates variance by mapping how the global minimum shifts in response to empirical noise perturbations, thus bypassing the need for matrix inversion. The approach adopted was as follows

At first the raw residual vectors $\{e_i\}_{i=A,B,C}$ corresponding to the optimal global fit of $\hat{\theta}$ are isolated for each peak. This is taken as "experimental noise". New synthetic noise vectors $\{e_i^*\}_{i=A,B,C}$ are then generated from this experimental noise by randomly picking their values and ascribing them to new positions in the new synthetic error vector (multiple occurrances of the experimental noise also being a possibility). New synthetic datasets are then constructed as $M_{syn,\,i} = M_i + e_i^*$, and the synthetic datasets is refitted. This procedure is then repeated multiple (in the present case, 50) independent times; the scattering statistics of the fitted parameters is then estimated from this ensemble of fits. The reported parameter uncertainties $\sigma_\theta$ are thus defined as the standard deviations of the resulting bootstrapped parameter distributions. To validate the uniqueness and to visualize parameter correlations, maps of the $\chi^2$ error surfaces are generated post optimization. These maps cover the objective function across a predefined $N \times N$ grid of input parameters, as set by the user's boundaries. Three specific planes are usually extracted to assess the integrity of the fit, $k_{ex}\ vs\ R_2$ , $k_{ex}\ vs\ M_i$, $k_{ex}\ vs\ \nu_i$ where $i = A, B, C$. The resulting $\chi^2$ plots are plotted logarithmically for visualization, to better enhance the dynamic range and clearly delineate the boundary of the confidence intervals.

**5.5 Execution.** All optimizations and executions were performed in Matlab®. The "lsqnonlin" solver was configured utilizing the Trust-Region-Reflective algorithm. To ensure the solver did not prematurely terminate in flat regions of the parameter space, strict termination tolerances were enforced for both the parameter step size (TolX $= 10^{-9}$) and the objective function gradient (TolFun $= 10^{-9}$). All figures shown were made with $N_{start} = 1000$ initial guesses, $B = 50$ bootstrap iterations and $50 \times 50$ sized $\chi^2$ grids. Examples of the ensuing pipeline (input and output data) are available upon request.

## Section 6: Experimental SSFP kinetics of DMA: Best fit parameters as a function of temperature for on- and off-resonance cases.

Figure S3 shows SSFP profiles of DMA (black) at three different temperatures recorded with peak B (33.1ppm) on-resonance, with best fit curves (red) overlaid. The fit parameters are given in Table S1. The corresponding error surface plots are provided in Section 6.1.

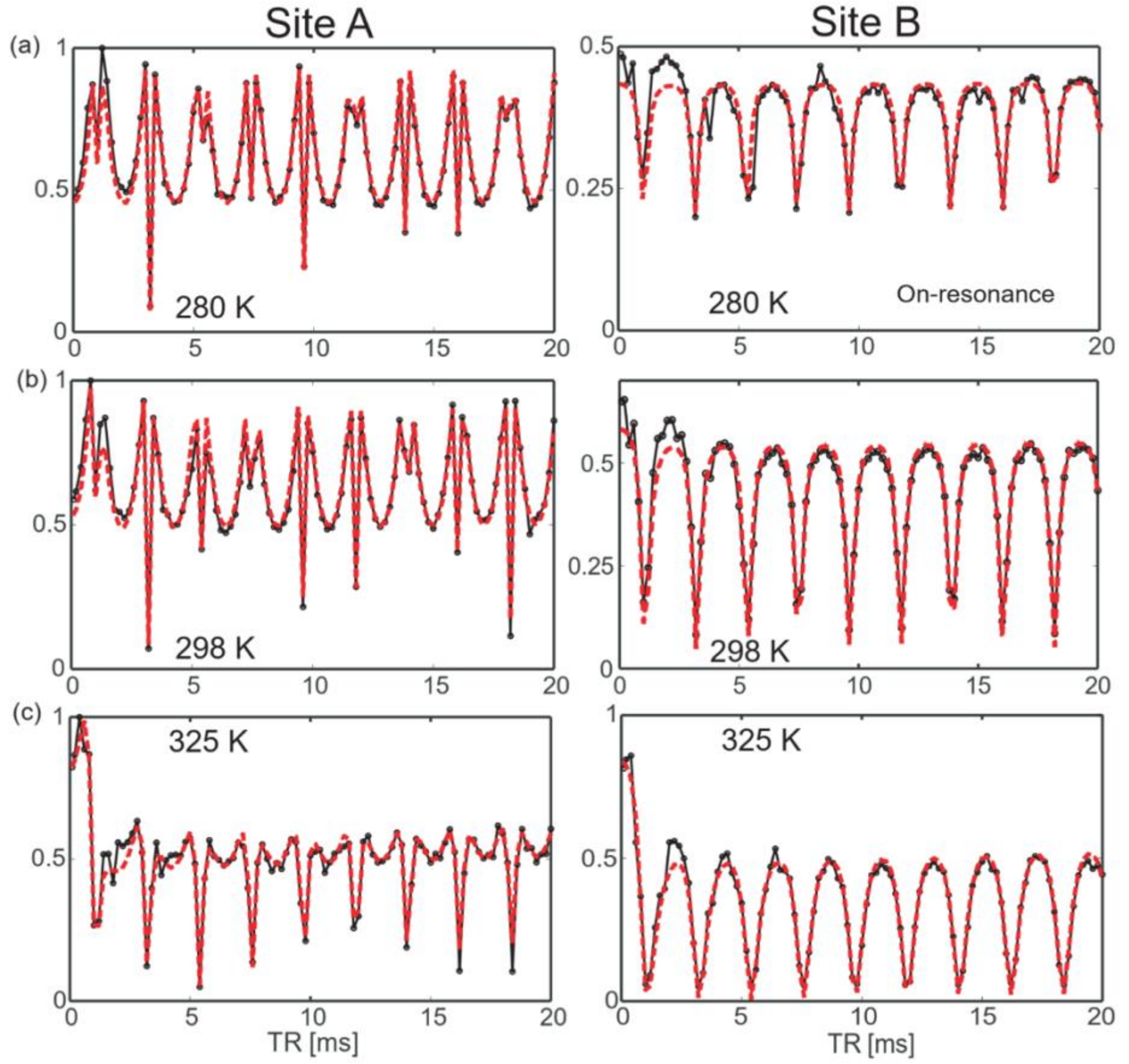

**Figure S3:** (a-c) On-resonance {[1]H}[13]C SSFP profiles (black) with best fit curves (red) for DMA at three different temperatures. Site B was on-resonance. Experiments were performed with TR ranging from 0.1 ms to 20.1 ms, varying in steps of 0.2 ms and a $10^o$ flip angle. The pulse sequence employed is shown in Figure 7 (main text), with $n{\cdot}TR$ = 72 s. The plots of site A and B are normalized globally to arrive at the best fit.

**Table S1:** Best-fit exchange parameters extracted from the data in Fig. S3; site B was on-resonance and it was assumed $R_{1A} = R_{1B} = 0.09\ s^{-1}$ (experimental measurement) and $R_{2A} = R_{2B}$.

| $Temp.$ [K] | $p_A$ | $p_B$ | $k_{ex}$ [$s^{-1}$] | $R_2$ [$s^{-1}$] | $\nu_A$ [Hz] | $\nu_B$ [Hz] | $\nu_A^{exp}$ [Hz] | $\nu_B^{exp}$ [Hz] |
|---|---|---|---|---|---|---|---|---|
| 280 | 0.51 $\pm$0.06 | 0.49 $\pm$ 0.05 | 0.2 $\pm$ 0.06 | 1.2 $\pm$ 0.3 | 469.8 $\pm$ 8.9 | 0 $\pm$ 2 | 469.9 | 0 |
| 298 | 0.48 $\pm$0.03 | 0.52 $\pm$0.03 | 2.2 $\pm$ 0.98 | 1.2 $\pm$ 0.4 | 467.4 $\pm$ 7.1 | 0.1 $\pm$ 2.1 | 467.7 | 0 |
| 325 | 0.5 $\pm$0.04 | 0.5 $\pm$0.04 | 19.4 $\pm$ 0.51 | 1.5 $\pm$ 0.24 | 462.5 $\pm$ 0.1 | $-0.7$ $\pm$ 2.6 | 463.4 | 0 |

### 6.1: $\chi^2$ plots associated to the best fit parameters tabulated in Table S1.

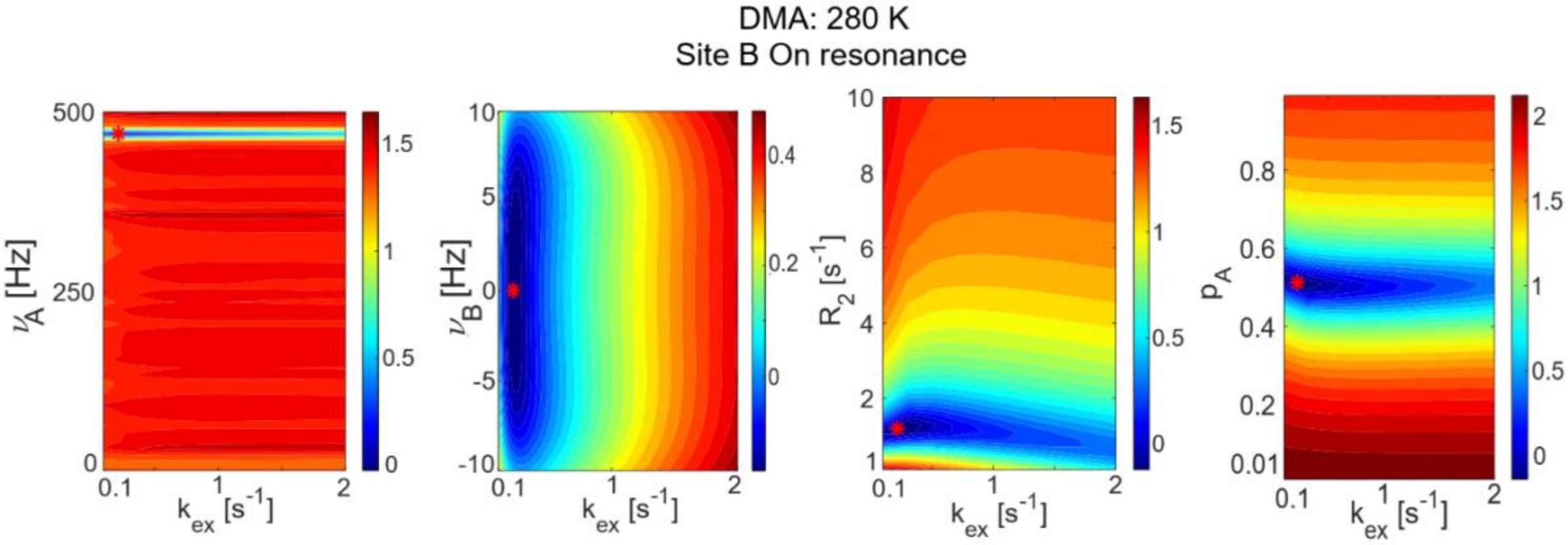

**Figure S4:** Error plots emerging from SSFP data on DMA at 280K, showing $\chi^2$ in the $k_{ex}$ vs $\nu_A$, $k_{ex}$ vs $\nu_B$, $k_{ex}$ vs $R_2$, and $k_{ex}$ vs population $p_A$ planes The boundaries considered for the fit were $p_A$= 0.01 to 0.99, $k_{ex} = 0.1$ to $10\ s^{-1}$; $R_2 = 1$ to $10\ s^{-1}$, $\nu_A = 0$ to 500 Hz; $\nu_B$ = 0 Hz. To appreciate the best fit values and the error surfaces, plots are shown only between $k_{ex} = 0.1$ and 2. The best fit parameters in all the plots are marked with a "*".

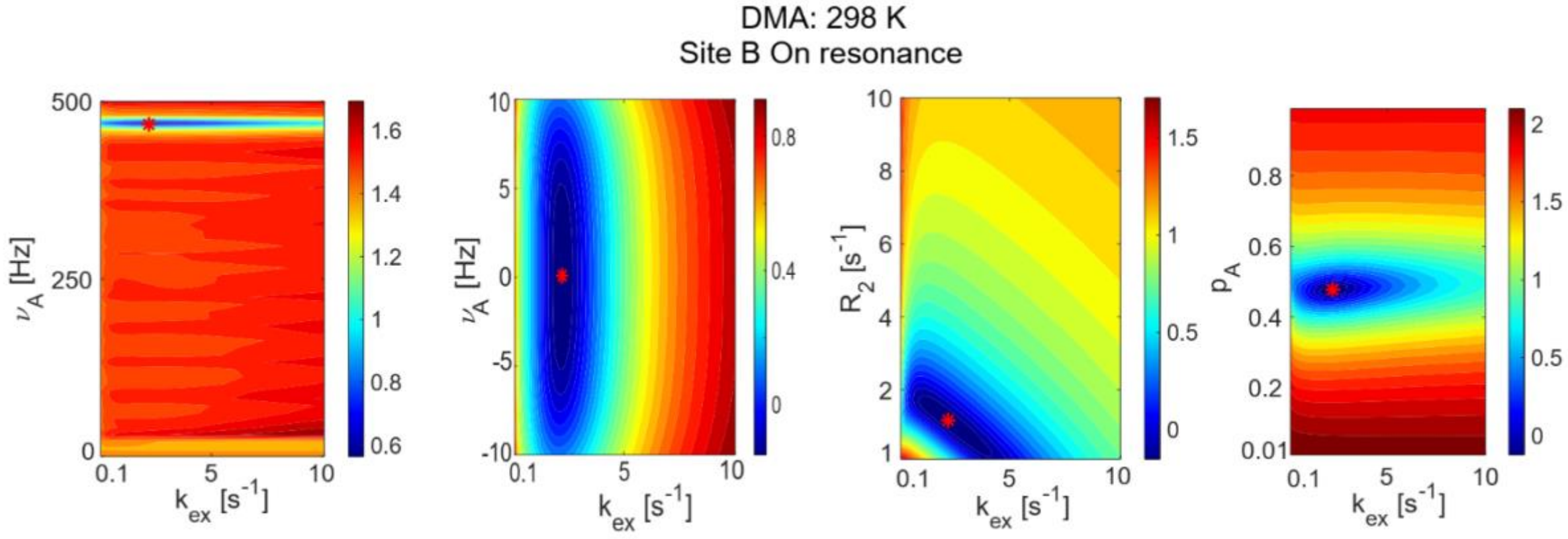

**Figure S5:** Idem as Figure S4, but at 298 K.

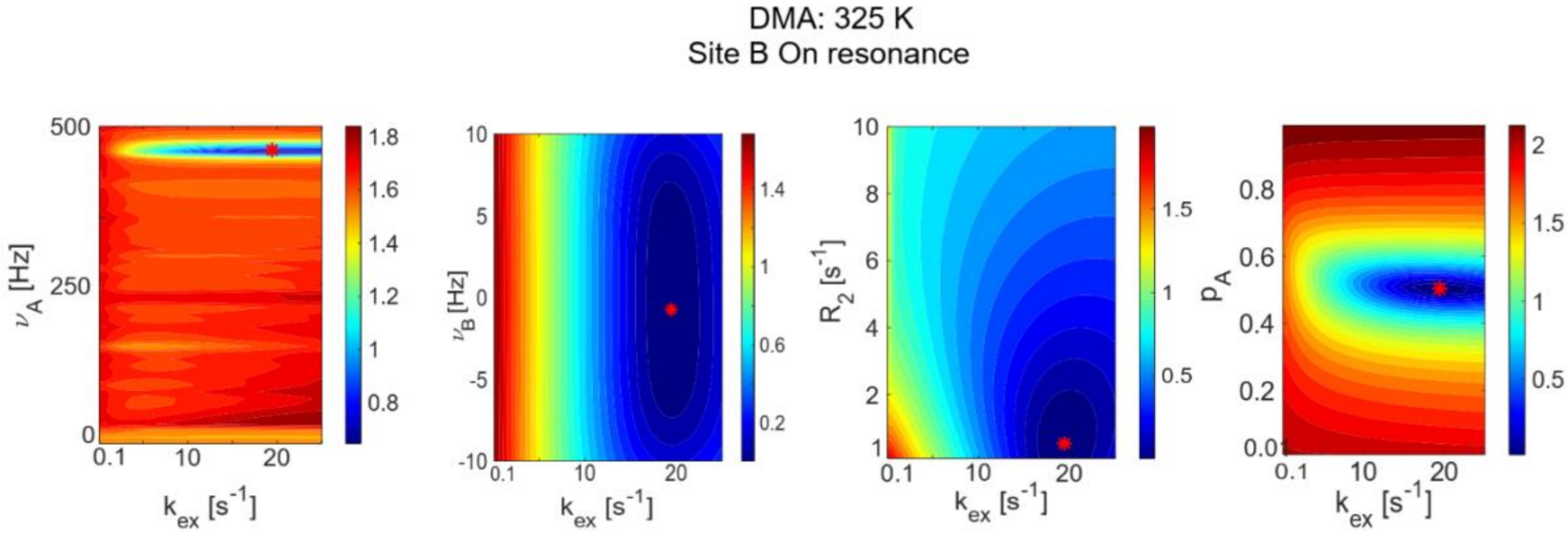

**Figure S6:** Idem as Figure S4, but with $k_{ex}$explored between 0.1 to 25 $s^{-1}$ and at 325 K.

**Section 6.2: : $\chi^2$ plots associated to the best fit parameters tabulated in Table I (Main Text).**

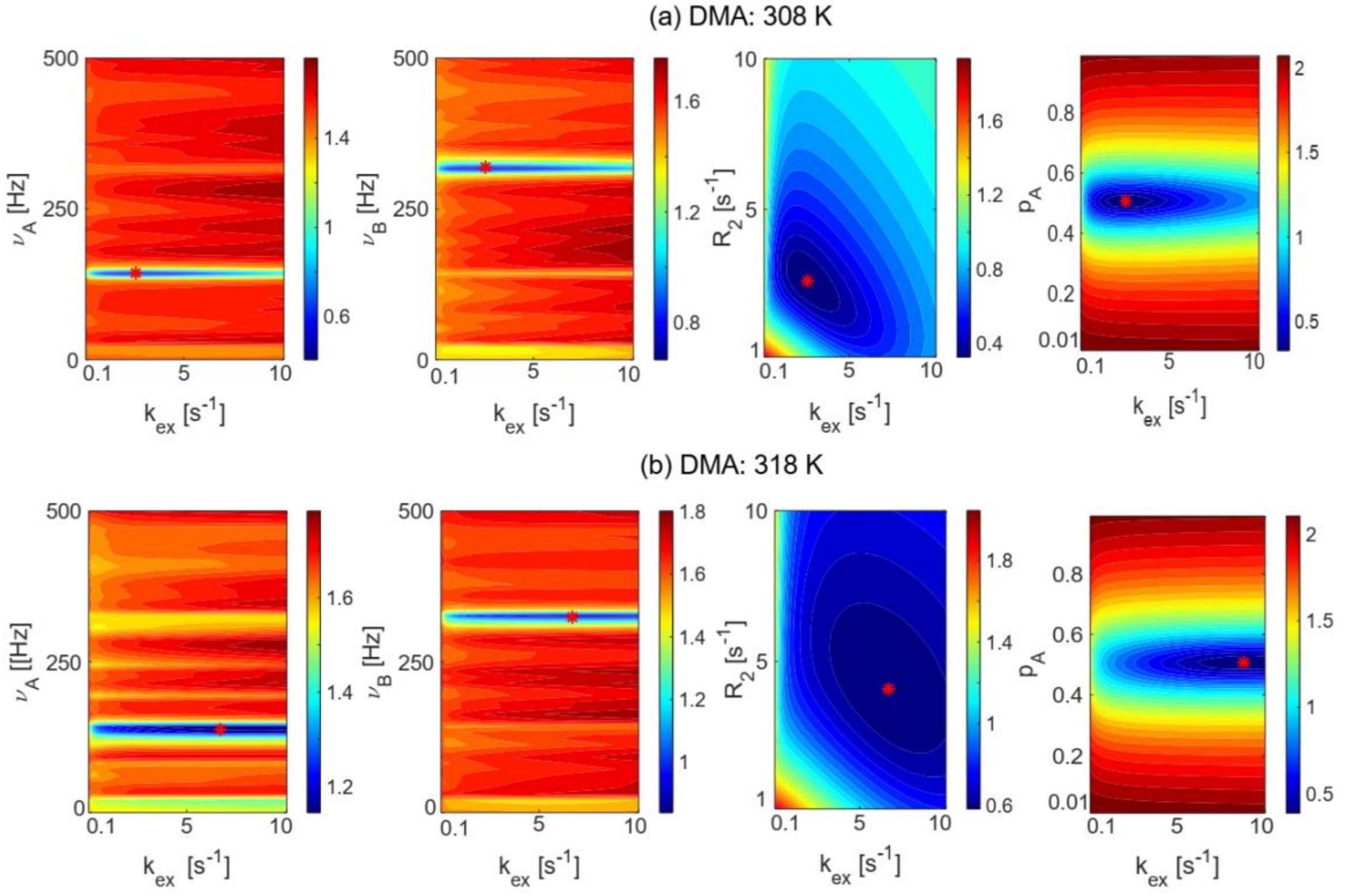

**Figure S7.** Error plots arising from SSFP experiments on DMA performed at an arbitrary offset for (a) 308 K and (b) 318 K. $\chi^2$ plots are shown in $k_{ex}$ vs $\nu_A$, $k_{ex}$ vs $\nu_B$, $k_{ex}$ vs $R_2$ and $k_{ex}$ vs population $p_A$ planes. The boundaries considered for the fit are for $p_A, p_B$= 0.01 to 0.99, $k_{ex} = 0.1$ to 10 $s^{-1}$; $R_2 =$ 1 to 10 $s^{-1}$, $\nu_A = 0$ to 500 Hz; $\nu_B = 0$ to 500 Hz; $R_{1A} = R_{1B} = 0.09\ s^{-1}$. The best fit parameters in all plots are marked with a "*".

## Section 7: SSFP profiles of AcAc with best fit parameters and errors as a function of temperature, for off- and on-resonance cases.

Figure S8 shows SSFP profiles of AcAc (black) at three different temperatures recorded with peak B (24.2 ppm) on-resonance, with best fit curves (red) overlaid. The fit parameters are given in Table S2. The corresponding error surface plots are provided in Section 7.1.

**Table S2 :** Exchange parameters extracted from the fitting program for AcAc with "site B" (enol-form methyl) on-resonance with the carrier and $R_{2A} = R_{2B}$.

| $Temp.$ [K] | $p_A$ | $p_B$ | $k_{ex}$ [s$^{-1}$] | $R_2$ [s$^{-1}$] | $\nu_A$ [Hz] | $\nu_B$ [Hz] | $\nu_A^{exp}$ [Hz] | $\nu_B^{exp}$ [Hz] |
|---|---|---|---|---|---|---|---|---|
| 293 | 0.28 $\pm$0.03 | 0.72 $\pm$ 0.08 | 8 $\pm$ 0.6 | 0.7 $\pm$ 0.2 | 927.8 $\pm$ 0.3 | $-12.6$ $\pm$ 3.5 | 927.6 | 0 |
| 310 | 0.28 $\pm$0.01 | 0.72 $\pm$0.03 | 34.3 $\pm$ 1.87 | 1.54 $\pm$ 0.15 | 923.5 $\pm$ 17.7 | 2.7 $\pm$ 3.9 | 923.8 | 0 |

| 325 | 0.32 $\pm$0.05 | 0.68 $\pm$0.10 | 61.2 $\pm$ 5.9 | 1.8 $\pm$ 0.15 | 920.8 $\pm$ 8.9 | 3.7 $\pm$ 2.4 | 918.4 | 0 |
|---|---|---|---|---|---|---|---|---|

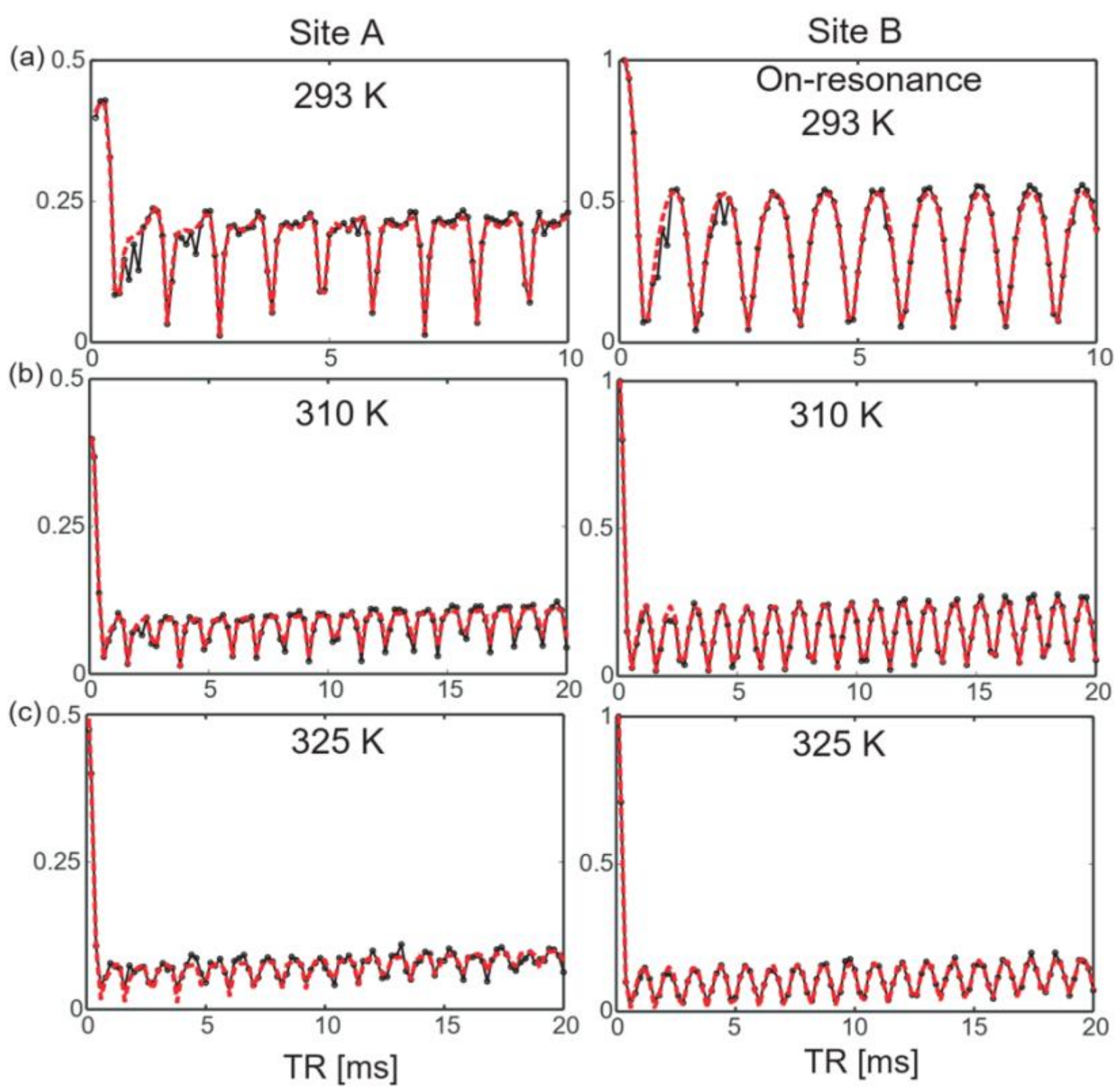

**Figure S8:** (a-c) On-resonance {[1]H}[13]C SSFP profiles (black) and best fit curves (red) for AcAc recorded at three different temperatures. The pulse sequence employed to record SSFP is shown in Figure 7 (main text). TR's were varied from 0.1 ms to 20.1 ms in steps of 0.2 ms; flip angle was $23^o$, and [13]C carrier was placed on the peak at 24.1 ppm (site B, enol). The plots of site A and B are normalized globally to arrive at the best fit parameters (Table S2).

**Section 7.1: $\chi^2$ plots associated to the best fit parameters tabulated in Table S2.**

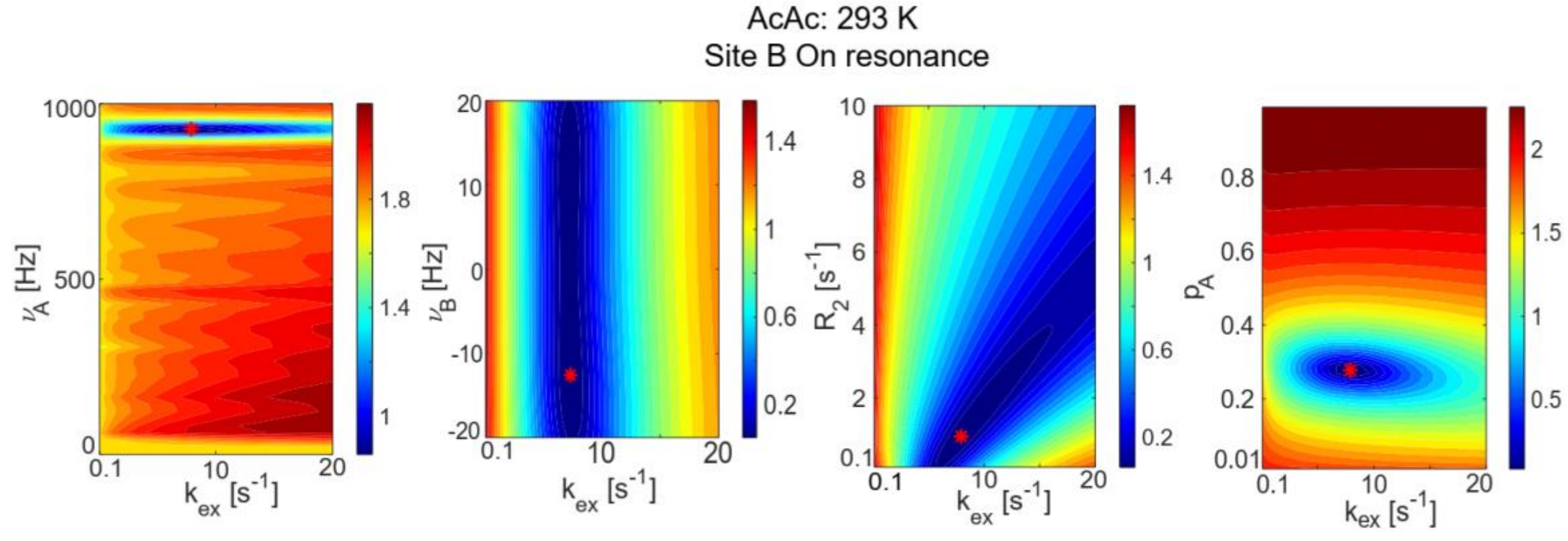

**Figure S9:** Error plots emerging from SSFP data on AcAc at 293 K, showing $\chi^2$ in the $k_{ex}$ vs $\nu_A$, $k_{ex}$ vs $\nu_B$, $k_{ex}$ vs $R_2$ and $k_{ex}$ vs population $p_A$ planes. The boundaries considered for the fit are for $p_A, p_B$= 0.01 to 0.99, $k_{ex} = 0.1$ to 20 $s^{-1}$; $R_2 = 1$ to 10 $s^{-1}$, $\nu_A = 0$ to 1000 Hz; $\nu_B$ = -20 to 20 Hz. The best fit parameters in all the plots are marked with a "*", and used experimentally measured $R_{1A} = R_{1B} = 0.09\ s^{-1}$ values.

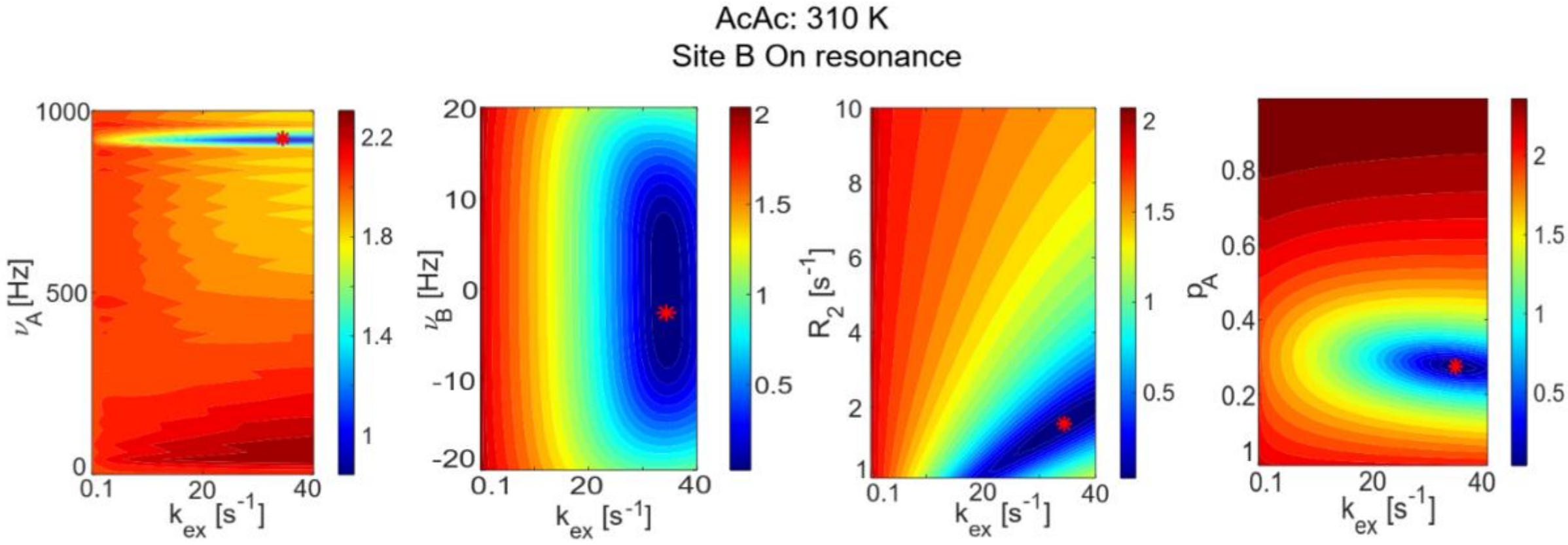

**Figure S10:** Idem as Figure S9 but at 310 K. Fits used measured $R_{1A} = R_{1B} = 0.077\ s^{-1}$ values and searched for $k_{ex}$ between 0.1 and 40 $s^{-1}$.

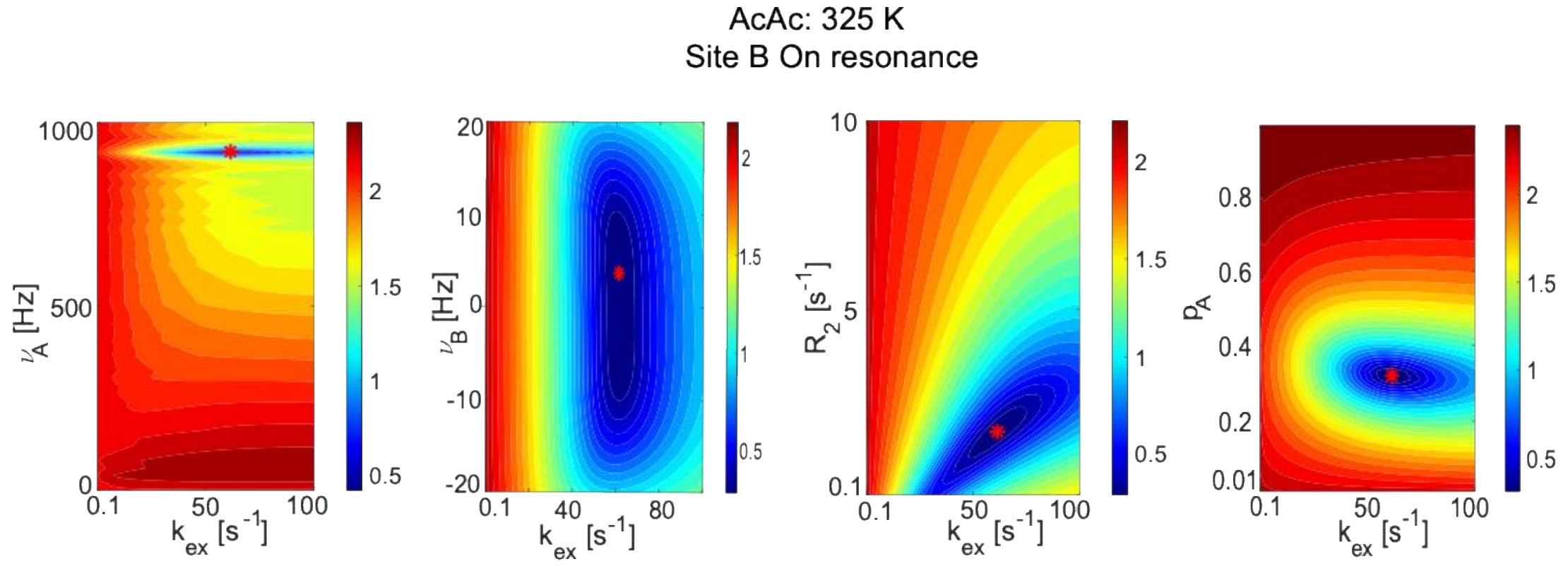

**Figure S11:** Idem as Figure S9 but at 325 K. Fits used measured $R_{1A} = R_{1B} = 0.069\ s^{-1}$, and searched for $k_{ex}$ between 0.1 and 100 $s^{-1}$.

## Section 7.2: $\chi^2$ plots associated to the AcAc best fit parameters tabulated in Table II (Main text).

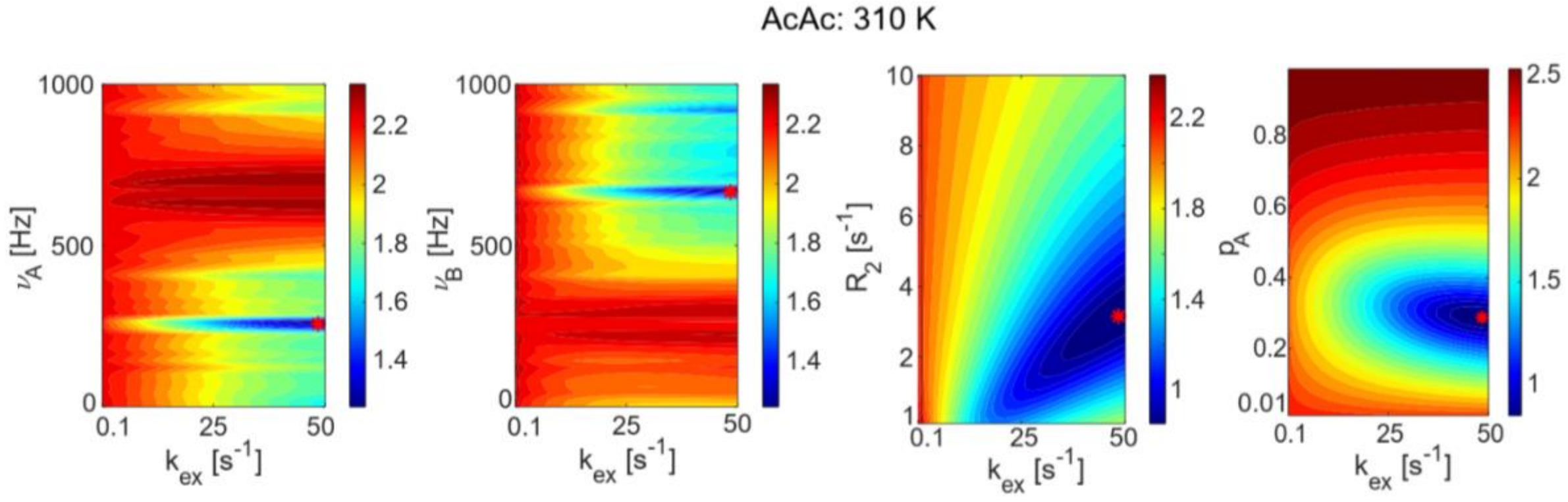

**Figure S12**. Errors in SSFP experiments performed on AcAc at an arbitrary offset at 310 K. Plots show $\chi^2$ in the $k_{ex}$ vs $\nu_A$, $k_{ex}$ vs $\nu_B$, $k_{ex}$ vs $R_2$, and $k_{ex}$ vs population $p_A$ planes. The boundaries considered for the fit are for $p_A, p_B$= 0.01 to 0.99, $k_{ex} = 0.1$ to 50 $s^{-1}$; $R_2 = 1$ to 10 $s^{-1}$, $\nu_A = 0$ to 1000 Hz; $\nu_B$ = 0 to 1000 Hz. The best fit parameters in all the plots are marked with a "*". To arrive at the best fit parameters, we have assumed $R_{1A} = R_{1B} = 0.069\ s^{-1}$.